\documentclass[prd,amsmath,superscriptaddress,notitlepage,twocolumn,eqsecnum,nofootinbib]{revtex4-2}
\pdfoutput=1
\usepackage[utf8]{inputenc}
\usepackage{times}
\usepackage{etoolbox}
\usepackage{titlesec}
\usepackage{graphicx}
\usepackage{float}
\usepackage{epstopdf,cancel}
\usepackage{epsf,latexsym,bbm,euscript}
\usepackage{amssymb}
\usepackage{mathtools}
\usepackage{xcolor}
\usepackage{csquotes}
\usepackage{soul}
\usepackage{mathtools}
\usepackage{braket}
\usepackage{comment}
\usepackage{tensor}
\usepackage{todonotes}
\usepackage[nice]{nicefrac}
\usepackage{enumerate}
\usepackage{placeins}
\usepackage{tikz}
\usetikzlibrary{positioning,shapes.geometric,decorations.markings,arrows,knots}
\usepackage{pgfplots}
\colorlet{bgcolor}{white}

\tikzset{
	overdraw/.style={preaction={draw,bgcolor,line width=#1}},
	overdraw/.default=5pt
}

\DeclareSymbolFont{matha}{OML}{txmi}{m}{it}
\DeclareMathSymbol{\varv}{\mathord}{matha}{118}

\makeatletter
\newcommand*{\defeq}{\mathrel{\rlap{%
			\raisebox{0.3ex}{$\m@th\cdot$}}%
		\raisebox{-0.3ex}{$\m@th\cdot$}}%
	=}
\newcommand*{\eqdef}{=\mathrel{\rlap{%
			\raisebox{0.3ex}{$\m@th\cdot$}}%
		\raisebox{-0.3ex}{$\m@th\cdot$}}%
}
\makeatother

\makeatletter
\g@addto@macro\bfseries{\boldmath}
\makeatother

\makeatletter
\def\thesubsection{\thesection.\arabic{subsection}}
\def\p@subsection{}
\makeatother

\titleformat{\section}[hang]
{\normalfont\bfseries\MakeUppercase}
{\thesection.}
{1em}
{\raggedright}[]

\titleformat{\subsection}[hang]
{\normalfont\bfseries}
{\thesubsection.}
{1em}
{\raggedright}[]

\definecolor{plotblue}{RGB}{0, 114, 178}
\definecolor{plotorange}{RGB}{255, 127, 14}
\definecolor{plotgreen}{RGB}{44, 160, 44}
\definecolor{plotpurple}{RGB}{128, 0, 128}

\usepackage{scalerel}
\usepackage{tikz}
\usetikzlibrary{svg.path}
\definecolor{orcidlogocol}{HTML}{A6CE39}
\tikzset{
	orcidlogo/.pic={
		\fill[orcidlogocol] svg{M256,128c0,70.7-57.3,128-128,128C57.3,256,0,198.7,0,128C0,57.3,57.3,0,128,0C198.7,0,256,57.3,256,128z};
		\fill[white] svg{M86.3,186.2H70.9V79.1h15.4v48.4V186.2z}
		svg{M108.9,79.1h41.6c39.6,0,57,28.3,57,53.6c0,27.5-21.5,53.6-56.8,53.6h-41.8V79.1z M124.3,172.4h24.5c34.9,0,42.9-26.5,42.9-39.7c0-21.5-13.7-39.7-43.7-39.7h-23.7V172.4z}
		svg{M88.7,56.8c0,5.5-4.5,10.1-10.1,10.1c-5.6,0-10.1-4.6-10.1-10.1c0-5.6,4.5-10.1,10.1-10.1C84.2,46.7,88.7,51.3,88.7,56.8z};
	}
}
\newcommand\orcidlink[1]{\href{https://orcid.org/#1}{\mbox{\scalerel*{
				\begin{tikzpicture}[yscale=-1,transform shape]
					\pic{orcidlogo};
				\end{tikzpicture}
			}{X}}}}

\usepackage{url,hyperref}
\hypersetup{colorlinks,linkcolor={blue!55!black},citecolor={red!50!black},urlcolor={blue!45!black},breaklinks=true}

\newcommand{\C}{\mathcal{C}}
\renewcommand{\hom}{\mathrm{hom}}
\renewcommand{\k}{\Bbbk}
\newcommand{\rmi}{\mathrm{i}}
\newcommand{\rme}{\mathrm{e}}
\newcommand{\T}{\mathrm{T}}
\newcommand{\YL}{\mathrm{YL}}
\newcommand{\SU}{\mathrm{SU}}
\newcommand{\qdi}[1]{[#1]_q}
\newcommand{\qddi}[1]{[#1]_{q'}}
\newcommand{\sixj}[6]{\begin{Bmatrix} #1 & #2 & #3 \\ #4 & #5 & #6 \end{Bmatrix}}

\usepackage{cleveref}

\begin{document}
	
\title{Categorical Quantum Volume Operator}

\author{Alexander Hahn\orcidlink{0000-0002-4152-9854}}
\email{alexander.hahn@students.mq.edu.au}
\affiliation{School of Mathematical and Physical Sciences, Macquarie University, Sydney, New South Wales 2109, Australia}

\author{Sebastian Murk\orcidlink{0000-0001-7296-0420}}
\email{sebastian.murk@oist.jp}
\affiliation{Quantum Gravity Unit, Okinawa Institute of Science and Technology, 1919-1 Tancha, Onna-son, Okinawa 904-0495, Japan}
	
\author{Sukhbinder Singh}
\email{qsukhi@gmail.com}
\affiliation{Multiverse Computing, Spadina Ave., Toronto, ON M5T 2C2, Canada}
	
\author{Gavin K. Brennen\orcidlink{0000-0002-6019-966X}}
\email{gavin.brennen@mq.edu.au}
\affiliation{School of Mathematical and Physical Sciences, Macquarie University, Sydney, New South Wales 2109, Australia}
	
\begin{abstract}
	We present a generalization of the quantum volume operator quantifying the volume in curved three-dimensional discrete geometries. In its standard form, the quantum volume operator is constructed from tetrahedra whose faces are endowed with irreducible representations of $\SU(2)$. Here, we show two equivalent constructions that allow general objects in fusion categories as degrees of freedom. First, we compute the volume operator for ribbon fusion categories. This includes the important class of modular tensor categories (such as quantum doubles), which are the building blocks of anyon models. Second, we further generalize the volume operator to spherical fusion categories by relaxing the categorical analog of the closure constraint (known as tetrahedral symmetry). In both cases, we obtain a volume operator that is Hermitian, provided that the input category is unitary. As an illustrative example, we consider the case of $\SU(2)_k$ and show that the standard $\SU(2)$ volume operator is recovered in the limit $k\rightarrow\infty$.
\end{abstract}
	
\maketitle
\tableofcontents
\section{Introduction} \label{sec:introduction}
In the search for an all-encompassing theory of quantum gravity that successfully reconciles general relativity and quantum field theory, background-independent theories such as loop quantum gravity (LQG)~\cite{Thiemann2007,Rovelli2014,Bilson-Thompson2024} and causal dynamical triangulations~\cite{Ambjorn2012,Loll2020} are widely regarded as the main contenders to string theory/M-theory. 
Arguably the most salient
property of background-independent theories is the prediction of a granular structure of space in which lengths, areas, and volumes are fundamentally discrete~\cite{RovelliSmolin1995,Loll1996,AshtekarLewandowski1997a,AshtekarLewandowski1997b,Thiemann1998,BrunnemannRideout2008a,BrunnemannRideout2008b}\@.
In the context of LQG, the discrete structure of spacetime has been instrumental in derivations of thermodynamic properties of black holes with possible observational consequences~\cite{Rovelli1996,AshtekarEA1998,Meissner2004,Perez2017}\@. It has also been proposed that a granular quantum spacetime might have observable imprints on the power spectrum of the cosmic microwave background radiation~\cite{AgulloEA2012,AgulloCorichi2014}\@. 

Due to the absence of a fixed background spacetime structure, states in background-independent theories are described by graphs, i.e.\ a collection of nodes (also referred to as ``vertices'') that are connected by weighted and in general directed links (also referred to as ``edges'').\footnote{In this article, we describe graphs using the language of string diagrams, see Sec.~\ref{subsec:string_diagrams}.}
The dual of the graph has a geometric interpretation according to which each $\varv$-valent node (i.e.\ a node with $\varv$ links attached to it) is associated with a $\varv$-polyhedron representing an elementary building block of the quantum spacetime geometry. 
Matter fields can only exist where this quantum geometry is excited, namely on the nodes (corresponding to fermionic degrees of freedom) and on the links (corresponding to bosonic degrees of freedom). Four-valent graphs whose duals are described by tetrahedra play a special role: as the smallest polyhedron with a nonzero volume they allow us to model the simplest physically relevant settings. The quantum volume operator is a fundamental observable in LQG. It was initially introduced for quantum spacetime states described by graphs whose edges carry representations of the group $\SU(2)$ and whose vertices carry linear maps that intertwine the representations on the incident edges.
Our aim in this article is to generalize the notion of the quantum volume operator to a large class of fusion categories beyond the representation category of $\SU(2)$.

The ``categorification'' of spin models has generated significant activity recently within the condensed matter and quantum information communities. The Levin--Wen string net model~\cite{LW04} is a prime example where a Hamiltonian constraint on a lattice with quantum degrees of freedom on the edges --- described by objects of a category $\C$ --- is constructed such that the ground states exhibit topological order through the mechanism of string net condensation. It was shown~\cite{Kadar2010} that the string net space for a category $\C$ is equal to the state space of the 3D Turaev--Viro topological quantum field theory (TQFT) for $\C$, or a 4D TQFT in the case of Walker--Wang models~\cite{Walker2012}\@. A microscopic construction of the Hamiltonian for general unitary fusion categories is given in Refs.~\cite{HW20,Lin2021} and has been further generalized to pivotal fusion categories in Ref.~\cite{Runkel2020}\@.

The string net model was motivated by categorifying lattice gauge theory where the gauge degrees of freedom located on the edges --- usually labeled by irreducible representations of a (gauge) group --- are supplanted by objects in a more general category. Other well-known examples of categorification are anyonic spin chains~\cite{AnyonicChains}, which generalize the quantum Heisenberg spin chain to anyonic degrees of freedom, such as the Golden Chain model of Fibonacci Anyons (objects of the category $SU(2)_3$). (Unlike String-net models which categorify gauge degrees of freedom, anyonic spin chains are obtained by categorifying the physical degrees of freedoms, namely, the lattice of spin $1/2$s in the Heisenberg model.) Similar and more general anyonic lattice models have since been shown to realize rich phase diagrams, see e.g., Refs.~\cite{SinghAnyons, AyeniAnyons}\@.
 
Motivated by these developments, one could ask whether the volume operator --- originally derived with isotropy of space in mind using irreducible representations from the group $\SU(2)$ --- could be generalized using objects from a category. Indeed, categorical extensions of loop quantum gravity are implied by the presence of a cosmological constant~\cite{DG14}\@. For 3D spacetime, one can write general relativity as a Chern--Simons gauge theory (see e.g. \cite{Bilson-Thompson2024}). The spin-foam approach to 3D gravity is obtained from combinatorial optimization of Chern--Simons theory with the quantum group $\mathcal{D}(U_q(\mathfrak{su}(2)))$, the Drinfeld double of $U_q(\mathfrak{su}(2))$ which is the $q$-deformation of the gauge group $\SU(2)$. In a 3D metric with Euclidean signature and with a positive cosmological constant $\Lambda$, the deformation parameter for the Chern--Simons theory is $q=\rme^{\rmi\ell_p/R}$, where $\ell_p$ denotes the Planck length and $R^{-1}=\sqrt{\Lambda}$ the cosmological radius. The irreducible representations of $\mathcal{D}(U_q(\mathfrak{su}(2)))$, also known as objects from the category $\C=\SU(2)_k$, are non-negative half-integers with an upper bound $k/2$ where the deformation is related to $k$ by $q=\rme^{\rmi\frac{2\pi}{k+2}}$. 

Earlier work~\cite{Smolinqdeformed, Smolinqdeformed1} computed a quantum volume for trivalent spin networks labeled by the objects in $\SU(2)_k$. However, such graphs have zero volume in the undeformed $k\rightarrow \infty$ limit and therefore do not correspond to the usual notion of volume in the classical limit with a classical symmetry group and labels with large spins. More recently, Livine showed~\cite{Livine2017} how to compute the action of the volume operator on an elementary tetrahedron in a Ponzano--Regge state sum model of 3D gravity. In that formulation, introduced in~\cite{LaurentFreidel_1999}, one defines the partition function from the action for a 2+1 dimensional BF theory with a cosmological term, where the three-dimensional manifold is triangulated by tetrahedra whose edges are labeled by objects from $\SU(2)_k$. The volume is computed by taking the derivative of the log of the Tureav-Viro transition amplitude, i.e. the partition function of the BF theory with a factor of $\rmi$ in the action, with respect to $-\rmi\Lambda$, and then setting $\Lambda\rightarrow 0$. It is important to note that Regge geometry is not the same as spin-network geometry since in the latter the shapes of faces of glued polyhedra do not have to match \cite{Charles2016}.

The main reason for the interest in categorical formulations and extensions of LQG is to circumvent the infrared divergences that appear using $\SU(2)$ with an infinite number of irreducible representations \cite{Rovelli_2011}. Additionally, other authors have recently pursued a categorical description of LQG for various reasons. Dittrich and Geiller~\cite{Dittrich2017} formulated a categorical description of the LQG kinematics, which clarifies the discreteness and finiteness of several quantum geometric operators~\cite{Dittrich2017}\@. However, they derive a categorical area operator, but not a volume operator. In a related work, Ref.~\cite{Dittrich2017a}, Dittrich speculates about defining grasping operators based on fusion category $F$-moves to define a categorical volume operator, which is closely related to the approach that we have developed in the present paper. Also, Ref.~\cite{Delcamp2017} shows how the categorical description of the kinematical Hilbert space for LQG provides a simple explanation of large-scale Gau{\ss} constraint violations, which is not apparent in the standard spin-network picture.

In this work, we compute the volume operator for an elementary four-valent node of a spin-network state, which has a quantum tetrahedron as its dual. We do this for the category $\SU(2)_k$ and several others. Our generalization of the $\SU(2)$ volume operator does not proceed from the quantization of classical gravity in the presence of a cosmological constant. Instead, it is based on preserving the categorical structure of the standard derivation~\cite[Ch.~7.5]{Rovelli2014} of the $\SU(2)$ volume operator. Clarifying the precise relation between the categorical SU(2)$_k$ volume operator that we derive here and $q$-deformed LQG, particularly how (if at all) our deformation of the volume operator relates to the cosmological constant, is beyond the scope of this work and remains an interesting question to explore in future research. Ultimately, our derivation of a categorical volume operator is motivated by the more general notion of categorification in quantum many-body physics, as outlined above. Nonetheless, our categorical volume operator fulfills a crucial constraint, namely, it reduces to the $\SU(2)$ volume operator in the limit $k \rightarrow \infty$, which previously known derivations~\cite{Smolinqdeformed, Smolinqdeformed1} have failed to demonstrate. Thus, we believe that our categorical volume operator is a viable candidate for further exploration as a geometric operator for LQG.

While we have motivated our volume operator mainly for applications in quantum gravity, the categorical volume operator may also prove useful for describing certain anyonic many-body systems. As outlined in Sec.~\ref{sec:chaktabarti}, the usual SU(2) volume operator of LQG was already derived by Chakrabarti to introduce a new quantum number for labeling symmetric (permutation-invariant) states of three spins~\cite{Chakrabarti1964}\@. Analogously, the categorical volume operator furnishes a quantum number to describe symmetric states of three  anyons. For example, it would be interesting to study how the Hilbert space of a chain of $n > 3$ anyons decomposes into quantum numbers under the action of the total volume operator, i.e., the sum of local volumes. These quantum numbers might yield a convenient and more compact representation of many-body anyonic states with certain spatial symmetries.
   	
The remainder of this article is organized as follows: In Sec.~\ref{sec:vol_op}, we briefly review the construction of the volume operator in LQG. In Sec.~\ref{sec:Qvol}, we provide a derivation for the generalization of the volume operator based on the properties of (spherical) fusion categories and demonstrate that this generalized version is Hermitian in the unitary case. In Sec.~\ref{sec:recovery}, we show that the generalized $\SU(2)_k$ volume operator reduces to the $\SU(2)$ volume operator in the limit $k\to\infty$. In Sec.~\ref{sec:spectrum}, we investigate the spectral properties of the $\SU(2)_k$ volume operator. Lastly, in Sec.~\ref{sec:discussion_conclusions}, we summarize our results, discuss their physical implications, and outline possible avenues for applications and directions for future research in this domain. The most relevant concepts and definitions from category theory that are used throughout this manuscript are reviewed in the \hyperref[sec:category_theory]{Appendix}\@.

\section{The Volume Operator in Loop Quantum Gravity} \label{sec:vol_op}
	
\subsection{Spin networks and graph states}
In LQG, a spin network state is defined on an oriented graph $\Gamma$. Each link $\ell$ has a direction and is labeled by a half-integer $j_\ell \in \mathbb{N} \frac{1}{2}$ spanning a $2 j_\ell + 1$ dimensional $\SU(2)$ irreducible representation space $\mathcal{H}_{j_\ell}$. Each node $\kappa$ of the graph carries a so-called intertwiner $\ket{\iota_\kappa}$. The intertwiner $\ket{\iota_\kappa}$ is a vector in the tensor representation $\mathcal{K}_\varv = \mathrm{Inv} \left( \mathcal{H}_{j_1} \otimes \mathcal{H}_{j_2} \otimes \cdots \mathcal{H}_{j_\varv} \right)$, which forms the kinematical Hilbert space. As alluded to in the introduction, the subscript $\varv$ labels the valence of the node. Furthermore, $\ket{\iota_\kappa}$ is invariant under the action of $\SU(2)$. Therefore, a spin network can be written as the triple $\ket{\Gamma, j_\ell, \iota_\kappa}$ defined by the tensor product of intertwiners at all nodes, i.e.\ $\ket{\Gamma, j_\ell, \iota_\kappa} = \otimes_\kappa \ket{\iota_\kappa}$.

\begin{figure}
	\centering
	\includegraphics{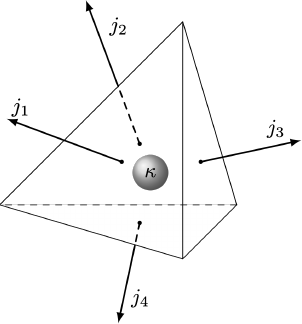}
	\caption{Schematic illustration of a quantum tetrahedron as the dual of a graph state. The node $\kappa$ is located in the center of the tetrahedron and carries a four-valent intertwiner $\ket{\iota_\kappa}$. A link $\ell$ in the graph state connected to this vertex pierces one of the tetrahedron's faces. It carries a half-integer spin $j_\ell\in\mathbb{N}\frac{1}{2}$ representing the area of the polyhedron face it pierces. More precisely, the quantum numbers $j_\ell$ are associated with the normal vectors $\vec{L}_\ell$ of the tetrahedron faces via the eigenvalue relation $\vec{L}_\ell^2\ket{j_\ell}=j_\ell(j_\ell+1)\ket{j_\ell}$.}
	\label{fig:Tetrahedron} 
\end{figure}

Euclidean polyhedra are the fundamental building blocks of arbitrary curved three-dimensional discrete geometries. In the simplest physically relevant setting, curved geometries are constructed from tetrahedra. In the dual (geometric) space, a four-valent intertwiner 
\begin{align}
	\ket{\iota_\kappa} = \sum\limits_{m_1 \cdots m_4} w^{m_1 \cdots m_4} \ket{j_1,m_1} \otimes \cdots \otimes \ket{j_4,m_4} ,	
\end{align} 
where $w^{m_1 \cdots m_4}$ is proportional to the Wigner $4jm$ symbol, is associated with a tetrahedron such that the node $\kappa$ is located inside of the tetrahedron with each of the four links $j_\ell$ connected to $\kappa$ piercing one of the tetrahedron faces as indicated in Fig.~\ref{fig:Tetrahedron}\@. It is always possible to decompose states described by nodes with valence $\varv \geqslant 4$ into a sum over states of trivalent (i.e.\ $\varv = 3$) intertwiners:
\begin{align}
	\begin{array}{lll}
		\begin{pmatrix}
			j_1 & j_2 & j_3 & j_4 \\
			m_1 & m_2 & m_3 & m_4
		\end{pmatrix}^{(j)}
		&=&
		\displaystyle{\sum_{m=-j}^j} (-1)^{j-m}	\begin{pmatrix}
			j_1 & j_2 & j \\
			m_1 & m_2 & m
		\end{pmatrix}
	\\
 &&	\begin{pmatrix}
			j & j_3 & j_4 \\
			-m & m_3 & m_4
		\end{pmatrix}
		,
	\end{array}
\end{align}
i.e.\ the Wigner $4jm$ symbol is decomposed into two Wigner $3jm$ symbols~\cite{MartinDussaud2019}\@. Diagrammatically, this can be represented as
\begin{equation*}
	\begin{tikzpicture}[baseline=(current bounding box.center)]
		\node (j1) at (-0.8,1) {\small$j_1,m_1$};
		\node (j2) at (-0.8,-1) {\small$j_2,m_2$};
		\node (j3) at (0.8,-1) {\small$j_3,m_3$};
		\node (j4) at (0.8,1) {\small$j_4,m_4$};
		\draw (j1) -- (j3);
		\draw (j2) -- (j4);
		\node (j) at (0.4,0) {\small $j$};
	\end{tikzpicture}
	=\sqrt{2j+1}
    \hspace*{-3mm}
	\begin{tikzpicture}[baseline=(current bounding box.center)]
		\node (j1) at (-1.5,1) {\small$j_1,m_1$};
		\node (j2) at (-1.5,-1) {\small$j_2,m_2$};
		\node (j3) at (1.5,-1) {\small$j_3,m_3$};
		\node (j4) at (1.5,1) {\small$j_4,m_4$};
		\draw (j1) -- (-0.7,0) -- (j2);
		\draw (j3) -- (0.7,0) -- (j4);
		\draw (-0.7,0) to node[above] {\small$j$} (0.7,0);
	\end{tikzpicture}
	\hspace*{-3mm} ,
\end{equation*}
where the virtual spin $j$ obeys the standard angular momentum coupling rules.

 \subsection{Quantum tetrahedron} \label{subsec:quantumtetrahedron}

In addition, intertwiners satisfy the closure constraint $\sum\limits_{\ell=1}^4 \vec{L}_\ell \ket{\iota_\kappa} = 0$, i.e.\ the sum of the normals $\vec{L}_\ell$ to the faces of a tetrahedron normalized to their area vanishes~\cite{Rovelli2014}:
\begin{equation}
	\vec{L}_1 + \vec{L}_2 + \vec{L}_3 + \vec{L}_4 = 0.
	\label{eq:closure_constraint}
\end{equation}
Thus any oriented triplet of edges gives the same result for the computation of the volume and it suffices to limit our considerations to the study of triplets of edges. Note also that for $\varv \geqslant 3$ a $\varv$-valent intertwiner possesses precisely $N_{\text{dof}} = \varv-3$ internal degrees of freedom, i.e.\ trivalent intertwiners are uniquely specified through their edges.
	
\subsection{Volume operator} \label{subsec:volumeoperator}
The action of the volume operator is given by~\cite{Rovelli2014}
\begin{equation}
	\hat{V} \ket{\iota_\kappa} = \frac{\sqrt{2}}{3} \left( 8 \pi G \hbar \gamma \right)^{3/2} \sqrt{\vert \hat{Q} \vert} \ket{\iota_\kappa} ,
	\label{eq:def:vol_op}
\end{equation}
where $\gamma$ denotes the Barbero--Immirzi parameter, and $G$ and $\hbar$ the gravitational and reduced Planck constant, respectively.\footnote{While the volume operator is technically $\hat{V} \propto \big(\hat{Q}^\dagger \hat{Q}\big)^{\nicefrac{1}{4}}$, where the dagger symbol $^\dagger$ indicates the adjoint, we take the liberty to refer to $\hat{Q}$ itself as the volume operator as well when there is no need (in terms of their physical relevance) to account for the various prefactors that appear in Eq. (\ref{eq:def:vol_op}).} The operator $\hat{Q}$ quantifies the volume of a single tetrahedron and is defined by
\begin{equation}
		\hat{Q} \defeq \vec{L}_1 \cdot \left( \vec{L}_2 \times \vec{L}_3 \right).
		\label{eq:def_volume_operator}
\end{equation}
The matrix elements $\tensor{Q}{_\kappa^{\kappa'}} = \braket{\iota_\kappa | \vec{L}_1 \cdot ( \vec{L}_2 \times \vec{L}_3 ) | \iota_{\kappa'}}$ of this operator are computed in Ref.~\cite{Rovelli2014}\@. We revisit the relevant parts of the derivation here, following Ref.~\cite{Rovelli2014} closely. First, we recall that due to the closure relation Eq.~\eqref{eq:closure_constraint}, any triple from $\{\vec{L}_1,\vec{L}_2,\vec{L}_3,\vec{L}_4\}$ can be used to derive $\hat{Q}$, and thus we can choose $\{\vec{L}_1,\vec{L}_2,\vec{L}_3\}$ [as in Eq.~\eqref{eq:def_volume_operator}] without loss of generality. Next, we define $L_{jk} \equiv L_j + L_k$ and compute
\begin{align}
	& \frac{1}{2}\left[(\vec{L}_1+\vec{L}_2)^2, \vec{L}_1\cdot\vec{L}_3\right] = \frac{1}{2}\left[ L_1^2 +L_2^2 + 2\vec{L}_1\cdot \vec{L}_2,\vec{L}_1\cdot \vec{L}_3\right] \nonumber \\
	& \qquad = \left[ \vec{L}_1\cdot \vec{L}_2, \vec{L}_1\cdot\vec{L}_3\right] = \sum_{j,k=1}^3 \left[ L_{1j} L_{2j}, L_{1k} L_{3k}\right] \nonumber \\
	& \qquad = \sum_{j,k,\ell=1}^3 \rmi \epsilon_{jk\ell} L_{1\ell} L_{2j} L_{3k} = \rmi \vec{L}_1 \cdot (\vec{L}_2\times\vec{L}_3), 
\end{align}
where $\epsilon_{jk\ell}$ denotes the Levi--Civita symbol. Therefore,
\begin{align}
	\begin{aligned}
		\tensor{Q}{_\kappa^{\kappa'}}
		=& -\frac{\rmi}{2}\bra{\iota_\kappa} [(\vec{L}_1+\vec{L}_2)^2, 	\vec{L}_1\cdot\vec{L}_3]\ket{\iota_{\kappa'}}\\
		=& -\frac{\rmi}{2} \big(\bra{\iota_\kappa} L_{12}^2(\vec{L}_1\cdot \vec{L}_3)\ket{\iota_{\kappa'}} \\
		&\qquad- \bra{\iota_\kappa} (\vec{L}_1\cdot\vec{L}_3)L_{12}^2\ket{\iota_{\kappa'}}\big)\\
		=&-\frac{\rmi}{2} (\alpha_\kappa - \alpha_{\kappa'}) \bra{\iota_\kappa} \vec{L}_1 \cdot \vec{L}_3 	\ket{\iota_{\kappa'}},
	\end{aligned}
\end{align}
where the last step follows from the Casimir property of $L_{12}^2$, that is
\begin{equation*}
	L_{12}^2\;
	\begin{tikzpicture}[baseline=(current bounding box.center)]
		\draw (0,-1) to node [left,midway] {$\kappa$} (0,0);
		\draw (0,0) to node [right,midway] {$j_2$} (0.75,0.75);
		\draw (0,0) to node [left,midway] {$j_1$} (-0.75,0.75);
	\end{tikzpicture}
	\;= \alpha_\kappa\;
	\begin{tikzpicture}[baseline=(current bounding box.center)]
		\draw (0,-1) to node [left,midway] {$\kappa$} (0,0);
		\draw (0,0) to node [right,midway] {$j_2$} (0.75,0.75);
		\draw (0,0) to node [left,midway] {$j_1$} (-0.75,0.75);
	\end{tikzpicture}\;.
\end{equation*}
For $\SU(2)$, we have $\alpha_\kappa = \kappa(\kappa+1)$. However, since one of our aims is to generalize this derivation, we do not make use of this specific relation here and proceed by referring to $\alpha_\kappa$ as the general eigenvalue of the quadratic Casimir operator instead. Using $(\vec{L}_1+\vec{L}_3)^2 = L_{13}^2 = L_1^2 + L_3^2 + 2 \vec{L}_1 \cdot \vec{L}_3$, we can compute
\begin{align}
	\tensor{Q}{_\kappa^{\kappa'}}
	&= - \frac{\rmi}{4} (\alpha_\kappa - \alpha_{\kappa'}) \bra{\iota_\kappa} L_{13}^2 - L_1^2 - L_3^2 \ket{\iota_{\kappa'}}\\
	&= - \frac{\rmi}{4} (\alpha_\kappa - \alpha_{\kappa'}) \big(\bra{\iota_\kappa} L_{13}^2 \ket{\iota_{\kappa'}} - (\alpha_{j_1} + \alpha_{j_3})\delta_{\kappa,\kappa'}\big), \nonumber
\end{align}
where we have made use of the fact that $L_1^2$ and $L_3^2$ are quadratic Casimir operators with eigenvalues $\alpha_{j_1},\alpha_{j_3}$, respectively. Note that the second term in the parentheses is zero as $(\alpha_\kappa - \alpha_{\kappa'})\delta_{\kappa,\kappa'}=0$. Therefore, we obtain
\begin{equation}
	\tensor{Q}{_\kappa^{\kappa'}} = - \frac{\rmi}{4} (\alpha_\kappa - \alpha_{\kappa'}) \bra{\iota_\kappa} L_{13}^2 \ket{\iota_{\kappa'}}
	\label{eq:volume_derivation_starting_point}
\end{equation}
as a final result. In Ref.~\cite{Rovelli2014}, the term $\bra{\iota_\kappa} L_{13}^2 \ket{\iota_{\kappa'}}$ is computed diagrammatically by inserting a resolution of the identity in terms of trivalent intertwiners, represented by a diagram of the form
\begin{equation}
	\begin{tikzpicture}[baseline=(current bounding box.center)]
		\draw[dashed] (0,0) arc (-40:40:1);
		\draw[rotate=180] (0,0) arc (40:-40:1);
		\draw (2,0) arc (-40:40:1);
		\draw[rotate=180] (-2,0) arc (40:-40:1);
		\draw (2,0) -- (1,-0.5);
		\draw (1,-0.5) -- (0,0);
		\draw (2,1.287) -- (1,1.8);
		\draw (1,1.8) -- (0,1.287);
		\node (kb) at (1,-0.8) {$\kappa'$};
		\node (k) at (1,2.1) {$\kappa$};
		\node (j1) at (-0.5,0.65) {$j_1$};
		\node (j4) at (2.5,0.65) {$j_4$};
		\node[draw,rectangle] (L13) at (1,0.65) {$L_{13}^2$};
		\draw (-0.225,0.65) -- (L13);
		\draw (1.775,0.65) -- (L13);
		\node (j3) at (1.6,1.15) {$j_3$};
		\node (j2) at (0.4,1.15) {$j_2$};
	\end{tikzpicture}\;.\label{eq:volume_diagram}
\end{equation}
The authors conclude that the matrix elements $\tensor{Q}{_\kappa^{\kappa'}}$ are nonzero only if $\kappa$ and $\kappa'$ differ by exactly one, and are given by~\cite{Rovelli2014}
\begin{equation}
	\begin{array}{lll}		
		c_\kappa &=& \tensor{Q}{_\kappa^{\kappa-1}}\\
		&=& - \frac{\rmi}{2} \left( \kappa(\kappa+1) - (\kappa-1)\kappa \right) \sqrt{2\kappa+1}\\
		&&\sqrt{2k-1} \sqrt{j_1(j_1+1)(2j_1+1)} \sqrt{j_3(j_3+1)(2j_3+1)} \\
		&&
		\begin{Bmatrix}
			j_1 & 1 & j_1 \\
			\kappa & j_2 & \kappa-1
		\end{Bmatrix}
		\begin{Bmatrix}
			j_3 & 1 & j_3 \\
			\kappa & j_4 & \kappa-1
		\end{Bmatrix} ,
	\end{array}
	\label{eq:QRS}
\end{equation}
where the curly brackets denote Wigner $6j$ symbols.\footnote{For a comprehensive derivation of the Wigner $6j$ symbol and possible decompositions, for instance into Wigner $3jm$ symbols, see Ref.~\cite{MartinDussaud2019}.} The resulting matrices are Hermitian and of the form
\begin{align}
	\tensor{Q}{_\kappa^{\kappa'}} 
	&=
	\begin{bmatrix}
		0 & c_1 & 0 & \cdots \\
		\overline{c_1} & 0 & c_2 & \\
		0 & \overline{c_2} & 0 & \\
		\vdots & & & \ddots
	\end{bmatrix} ,
\end{align}
which guarantees that the spectrum of the volume operator is non-degenerate. 
(The eigenvalues come in pairs $\pm\lambda_i$ with an additional zero for odd dimensions.)
	
\subsection{Chakrabarti's derivation of the volume operator}\label{sec:chaktabarti}	
Interestingly, the derivation of the matrix elements of $\hat{Q}$ was provided some sixty years ago by Chakrabarti~\cite{Chakrabarti1964}, who referred to it as an operator ``whose classical analog is a volume generated by the three angular momentum vectors". Chakrabarti was interested in constructing eigenstates, which reflect the permutation symmetry of three particles with angular momentum $\vec{L}_1,\vec{L}_2,\vec{L}_3$ and total angular momentum $\vec{L}_4=\sum_{k}\vec{L}_k$, which is the same as our closure constraint Eq.~\eqref{eq:closure_constraint}\@. In the basis $\{\ket{\iota_k}\}$, which is an eigenbasis of $(\vec{L}_1+\vec{L}_2)^2$ with eigenvalue $\kappa(\kappa+1)$, he showed that
\begin{align}
	\begin{aligned}
		\bra{\iota_\kappa'}\hat{Q}\ket{\iota_\kappa}=&-\frac{\rmi}{4} \big(\kappa(\kappa+1)-\kappa'(\kappa'+1)\big) \\
		&\sqrt{(2\kappa+1)(2\kappa'+1)} \\
		&\sum_{\ell\in\mathcal{L}_{23}}\ell(\ell+1)(2\ell+1) \\
		&\begin{Bmatrix}
			j_1 & j_2 & \kappa \\
			j_3 & j_4 & \ell
		\end{Bmatrix}
		\begin{Bmatrix}
			j_1 & j_2 & \kappa' \\
			j_3 & j_4 & \ell
		\end{Bmatrix},
	\end{aligned}
	\label{eq:QCh}
\end{align}
where $\mathcal{L}_{23} \defeq \{|j_2-j_3|,\dots,j_2+j_3\}\cap\{|j_1-j_4|,\dots,j_1+j_4\}$. This result follows by writing $\hat{Q}=\frac{\rmi}{4}[(\vec{L}_1+\vec{L}_2)^2,(\vec{L}_2+\vec{L}_3)^2]$, transforming to a basis in which $(\vec{L}_2+\vec{L}_3)^2$ is diagonal via the Wigner $6j$ symbols, evaluating the action of $(\vec{L}_2+\vec{L}_3)^2$ which returns the Casimir for the coupled angular momenta, and then transforming back. If instead one evaluates $(\vec{L}_2+\vec{L}_3)^2$ directly in the basis $\{\ket{\iota_k}\}$, one finds, as first shown in Ref.~\cite{LevyLeblond1965}, the same result for the matrix elements as in Eq.~\eqref{eq:QRS}, which is equivalent to Eq.~\eqref{eq:QCh}\@. The selection rule $\kappa' = \kappa \pm 1$ arises from the fact that the evaluation of $\vec{L}_2\cdot \vec{L}_3$ involves a rank-one operator $L_2$ in the coupled basis $\ket{\iota_\kappa}$, which implies a selection rule $|\kappa'-\kappa|=1$. 
	
\section{Categorification of the Volume Operator} \label{sec:Qvol}
In this section, we outline the framework for generalizing the quantum volume operator introduced in the previous section to a general class of categories. We briefly introduce the category theoretical concepts needed for our construction in the \hyperref[sec:category_theory]{Appendix}\@. For a more comprehensive introduction to category theory, we refer the reader to Refs.~\cite{Beer2018,Wolf2020,Etingof2015}\@. We will assume that $\C$ is a ribbon fusion category, sometimes also called a premodular category. By Tannaka duality~\cite{Joyal1990}, such categories are the representation categories of ribbon Hopf algebras. These are semisimple quasi-triangular Hopf algebras with additional constraints on the ribbon element.
	
Ribbon fusion categories form a broad class of physically relevant models. An important class of examples is modular tensor categories, which are the mathematical foundation for anyon models~\cite{Beer2018,Bonderson2007}\@. Note that one can canonically construct a modular tensor category from any spherical fusion category via the Drinfeld quantum double~\cite{Mueger2003}\@.
	
\subsection{String diagrams} \label{subsec:string_diagrams}
A fusion category is uniquely specified by its skeletal data, see e.g.\ Refs.~\cite{Bridgeman2020,Barter2022}\@. That is
\begin{enumerate}[(i)]
	\item a finite set $I=\{0,a,b,\dots\}$ of simple objects as well as their duals $0,a^*,b^*,\dots$, where $0$ labels the unit object;
	\item the fusion rules for all $a,b\in I$,
	\begin{equation*}
		a\otimes b = \sum_{c\in I} N_{ab}^{c} c,
	\end{equation*}
	together with the fusion spaces $\hom(a\otimes b,c)$ [which are vector spaces of dimension $N_{ab}^c$] represented by trivalent vertices
	\begin{equation*}
		\begin{tikzpicture}[decoration={markings,mark=at position 0.5 with {\arrow[scale=1,thick,>=stealth]{>}}},baseline=(current bounding box.center),rotate=180] 
			\draw[{postaction=decorate}] (0,0) to node [right,midway] {$c$} (0,-1);
			\draw[{postaction=decorate}] (0.75,0.75) to node [above left,midway] {$a$} (0,0);
			\draw[{postaction=decorate}] (-0.75,0.75) to node [above right,midway] {$b$} (0,0);
		\end{tikzpicture}\;,
	\end{equation*}
	and
	\item the $F$-symbols, which relate different fusion paths to each other via relations of the form
	\begin{equation}
		\begin{tikzpicture}[scale=1,baseline=(current bounding box.center), decoration={markings,mark=at position .6 with {\arrow[>=stealth]{>}}}]
			\node (u1) at (0,0) {$d$};
			\coordinate (v11) at (0,0.5);
			\coordinate (v12) at (-0.5,1);
			\node (x1) at (-1,1.5) {$a$};
			\node (y1) at (0,1.5) {$b$};
			\node (z1) at (1,1.5) {$c$};
				
			\draw[{postaction=decorate}] (u1) -- (v11);
			\draw[{postaction=decorate}] (v11) -- (z1);
			\draw[{postaction=decorate}] (v11) to node[below left] {$e$} (v12);
			\draw[{postaction=decorate}] (v12) -- (y1);
			\draw[{postaction=decorate}] (v12) -- (x1);
		\end{tikzpicture}
		=\sum_{f} \big(F_d^{abc}\big)_{ef}
		\begin{tikzpicture}[scale=1,baseline=(current bounding box.center), decoration={markings,mark=at position .6 with {\arrow[>=stealth]{>}}}]
			\node (u2) at (0,0) {$d$};
			\coordinate (v21) at (0,0.5);
			\coordinate (v22) at (0.5,1);
			\node (x2) at (-1,1.5) {$a$};
			\node (y2) at (0,1.5) {$b$};
			\node (z2) at (1,1.5) {$c$};
				
			\draw[{postaction=decorate}] (u2) -- (v21);
			\draw[{postaction=decorate}] (v21) -- (x2);
			\draw[{postaction=decorate}] (v21) to node[below right] {$f$} (v22);
			\draw[{postaction=decorate}] (v22) -- (y2);
			\draw[{postaction=decorate}] (v22) -- (z2);
		\end{tikzpicture}\;.\label{eq:F-symbols}
	\end{equation}		
\end{enumerate}
For notational simplicity, we work with multiplicity-free categories, i.e.\ $N_{ab}^c=1$ for all $a,b,c\in I$. However, this is not a necessary assumption and our construction can be made general by introducing an additional fusion label to each trivalent vertex. The $F$-symbols are the category-theoretic generalization of the Wigner $6j$ symbols.
	
For the fusion spaces in $\C$, we fix the standard basis
\begin{equation}
	\begin{tikzpicture}[scale=1.5,baseline=(current bounding box.center), decoration={markings,mark=at position .6 with {\arrow[>=stealth]{>}}}]
		\coordinate (ydown) at (0,0);
		\coordinate (yup) at (0,0.5);
		\coordinate (t12) at (-0.5,1);
		\node (x1) at (-1,1.5) {$a_1$};
		\node (x2) at (0,1.5) {$a_2$};
		\node (xm) at (1,1.5) {$a_m$};
		\draw[{postaction=decorate}] (ydown) to node[right] {$b$} (yup);
		\draw[{postaction=decorate}] (yup) -- (xm);
		\draw[{postaction=decorate}] (yup) to node[below left] {$b_2$} (t12);
		\draw[{postaction=decorate}] (t12) -- (x2);
		\draw[{postaction=decorate}] (t12) -- (x1);
		\node (xsn) at (1,-1) {$a'_n$};
		\draw[{postaction=decorate}] (xsn) -- (ydown);
		\node (xs1) at (-1,-1) {$a'_1$};
		\node (xs2) at (0,-1) {$a'_2$};
		\coordinate (t12s) at (-0.5,-0.5);
		\draw[{postaction=decorate}] (xs1) -- (t12s);
		\draw[{postaction=decorate}] (xs2) -- (t12s);
		\draw[{postaction=decorate}] (t12s) to node[above left] {$b'_2$} (ydown);
		\node (downdots) at (0.5,-1) {$\cdots$};
		\node (updots) at (0.5,1.5) {$\cdots$};
		\node[rotate=45] (ysdots) at (-0.15,-0.35) {$\cdots$};
		\node[rotate=-45] (ydots) at (-0.15,0.85) {$\cdots$};
	\end{tikzpicture}\label{eq:standard_basis}
\end{equation}
and choose the following normalization for the trivalent vertices:
\begin{equation}
	\left(\frac{d_c}{d_ad_b}\right)^{1/4}\;
	\begin{tikzpicture}[decoration={markings,mark=at position 0.5 with {\arrow[scale=1,thick,>=stealth]{>}}},baseline=(current bounding box.center)] 
		\draw[{postaction=decorate}] (0,-1) to node [left,midway] {$c$} (0,0);
		\draw[{postaction=decorate}] (0,0) to node [below right,midway] {$b$} (0.75,0.75);
		\draw[{postaction=decorate}] (0,0) to node [below left,midway] {$a$} (-0.75,0.75);
	\end{tikzpicture}\;,
	\quad
	\left(\frac{d_c}{d_ad_b}\right)^{1/4}\;
	\begin{tikzpicture}[decoration={markings,mark=at position 0.5 with {\arrow[scale=1,thick,>=stealth]{>}}},baseline=(current bounding box.center),rotate=180] 
		\draw[{postaction=decorate}] (0,0) to node [right,midway] {$c$} (0,-1);
		\draw[{postaction=decorate}] (0.75,0.75) to node [above left,midway] {$a$} (0,0);
		\draw[{postaction=decorate}] (-0.75,0.75) to node [above right,midway] {$b$} (0,0);
	\end{tikzpicture}\;,\label{eq:normalization_trivalent_vertices}
\end{equation}
where $d_a$ denotes the quantum dimension of the object $a\in\C$,
\begin{equation}
	d_a=
	\begin{tikzpicture}[decoration={markings,mark=at position .5 with {\arrow[scale=1,thick,>=stealth]{>}}},baseline=(current bounding box.center)]
		\draw[{postaction=decorate}] (0,-0.5) to node [left,midway] {$a$} (0,0.5);
		\draw (0,0.5) -- (0.5,1) -- (1,0.5);
		\draw[{postaction=decorate}] (1,0.5) to node [right,midway] {$a$} (1,-0.5);
		\draw (1,-0.5) -- (0.5,-1) -- (0,-0.5);
	\end{tikzpicture}
	\; , \label{eq:quantum_dimension}
\end{equation}
and $d_a=d_{a^*}$ for all objects $a\in\C$. This normalization convention is standard in the physics literature as it ensures isotopy invariance~\cite{Bonderson2007}\@. That is, the value of a string diagram remains invariant when the strings are continuously deformed keeping their open ends fixed in place. Using this normalization convention, we obtain the bigon relation
\begin{equation}
	\begin{tikzpicture}[decoration={markings,mark=at position .5 with {\arrow[scale=1,thick,>=stealth]{>}}},baseline=(current bounding box.center)]
		\draw[{postaction=decorate}] (0,-0.5) to node [left,midway] {$a$} (0,0.5);
		\draw (0,0.5) -- (0.5,1) -- (1,0.5);
		\draw[{postaction=decorate}] (0.5,1) to node [left,midway] {$c'$} (0.5,1.5);
		\draw[{postaction=decorate}] (1,-0.5) to node [right,midway] {$b$} (1,0.5);
		\draw (1,-0.5) -- (0.5,-1) -- (0,-0.5);
		\draw[{postaction=decorate}] (0.5,-1.5) to node [left,midway] {$c$} (0.5,-1);
	\end{tikzpicture}
	=\delta_{c,c'}\sqrt{\frac{d_a d_b}{d_c}}\quad
	\begin{tikzpicture}[decoration={markings,mark=at position .5 with {\arrow[scale=1,thick,>=stealth]{>}}},baseline=(current bounding box.center)]
		\draw[{postaction=decorate}] (0,-1.5) to node [right,midway] {$c$} (0,1.5);
	\end{tikzpicture}
	\; \label{eq:bigon_relation}
\end{equation}
and the completeness relation (resolution of the identity)
\begin{equation}
	\begin{tikzpicture}[decoration={markings,mark=at position .5 with {\arrow[scale=1,thick,>=stealth]{>}}},baseline=(current bounding box.center)]
		\draw[{postaction=decorate}] (0,0) to node[left,midway] {$a$} (0,2);
		\draw[{postaction=decorate}] (1,0) to node[right,midway] {$b$} (1,2);
	\end{tikzpicture}\;
	= \sum_{c} \sqrt{\frac{d_c}{d_a d_b}}\;
	\begin{tikzpicture}[decoration={markings,mark=at position .5 with {\arrow[scale=1,thick,>=stealth]{>}}},baseline=(current bounding box.center)]
		\draw[{postaction=decorate}] (-0.5,-1) to node[above left] {$a$} (0,-0.5);
		\draw[{postaction=decorate}] (0.5,-1) to node[above right] {$b$} (0,-0.5);
		\draw[{postaction=decorate}] (0,-0.5) to node[right] {$c$} (0,0.5);
		\draw[{postaction=decorate}] (0,0.5) to node[below left] {$a$} (-0.5,1);
		\draw[{postaction=decorate}] (0,0.5) to node[below right] {$b$} (0.5,1);
	\end{tikzpicture}\;.\label{eq:completeness_relation}
\end{equation}
By fixing the standard basis~\eqref{eq:standard_basis}, the $F$-symbols~\eqref{eq:F-symbols} can be represented as matrices with indices $e,f$. If these matrices are unitary, we call $\C$ a \emph{unitary} ribbon fusion category. In this case, all string diagrams satisfy mirror symmetry. That is, flipping a diagram along the horizontal axis corresponds to taking its complex conjugate with respect to the standard basis~\eqref{eq:standard_basis}\@. Therefore, for unitary $\C$, we have
\begin{equation*}
	\begin{tikzpicture}[scale=1,baseline=(current bounding box.center), decoration={markings,mark=at position .6 with {\arrow[>=stealth]{<}}},rotate=180]
		\node (u1) at (0,0) {$d$};
		\coordinate (v11) at (0,0.5);
		\coordinate (v12) at (-0.5,1);
		\node (x1) at (-1,1.5) {$c$};
		\node (y1) at (0,1.5) {$b$};
		\node (z1) at (1,1.5) {$a$};
			
		\draw[{postaction=decorate}] (u1) -- (v11);
		\draw[{postaction=decorate}] (v11) -- (z1);
		\draw[{postaction=decorate}] (v11) to node[above right] {$e$} (v12);
		\draw[{postaction=decorate}] (v12) -- (y1);
		\draw[{postaction=decorate}] (v12) -- (x1);
	\end{tikzpicture}
	=\sum_{f} \overline{\big(F_d^{abc}\big)}_{ef}
	\begin{tikzpicture}[scale=1,baseline=(current bounding box.center), decoration={markings,mark=at position .6 with {\arrow[>=stealth]{<}}},rotate=180]
		\node (u2) at (0,0) {$d$};
		\coordinate (v21) at (0,0.5);
		\coordinate (v22) at (0.5,1);
		\node (x2) at (-1,1.5) {$c$};
		\node (y2) at (0,1.5) {$b$};
		\node (z2) at (1,1.5) {$a$};
			
		\draw[{postaction=decorate}] (u2) -- (v21);
		\draw[{postaction=decorate}] (v21) -- (x2);
		\draw[{postaction=decorate}] (v21) to node[above left] {$f$} (v22);
		\draw[{postaction=decorate}] (v22) -- (y2);
		\draw[{postaction=decorate}] (v22) -- (z2);
	\end{tikzpicture}\;,
\end{equation*}
while in the general non-unitary case
\begin{equation*}
	\begin{tikzpicture}[scale=1,baseline=(current bounding box.center), decoration={markings,mark=at position .6 with {\arrow[>=stealth]{<}}},rotate=180]
		\node (u1) at (0,0) {$d$};
		\coordinate (v11) at (0,0.5);
		\coordinate (v12) at (-0.5,1);
		\node (x1) at (-1,1.5) {$c$};
		\node (y1) at (0,1.5) {$b$};
		\node (z1) at (1,1.5) {$a$};
			
		\draw[{postaction=decorate}] (u1) -- (v11);
		\draw[{postaction=decorate}] (v11) -- (z1);
		\draw[{postaction=decorate}] (v11) to node[above right] {$e$} (v12);
		\draw[{postaction=decorate}] (v12) -- (y1);
		\draw[{postaction=decorate}] (v12) -- (x1);
	\end{tikzpicture}
	=\sum_{f} \big(F_d^{abc}\big)_{ef}^\T
	\begin{tikzpicture}[scale=1,baseline=(current bounding box.center), decoration={markings,mark=at position .6 with {\arrow[>=stealth]{<}}},rotate=180]
		\node (u2) at (0,0) {$d$};
		\coordinate (v21) at (0,0.5);
		\coordinate (v22) at (0.5,1);
		\node (x2) at (-1,1.5) {$c$};
		\node (y2) at (0,1.5) {$b$};
		\node (z2) at (1,1.5) {$a$};
			
		\draw[{postaction=decorate}] (u2) -- (v21);
		\draw[{postaction=decorate}] (v21) -- (x2);
		\draw[{postaction=decorate}] (v21) to node[above left] {$f$} (v22);
		\draw[{postaction=decorate}] (v22) -- (y2);
		\draw[{postaction=decorate}] (v22) -- (z2);
	\end{tikzpicture}\;,
\end{equation*}
where the superscript $\T$ denotes the transpose with respect to the standard basis. In ribbon fusion categories, the $F$-symbols satisfy the tetrahedral symmetry relation~\cite{Turaev2017}
\begin{equation}
	\begin{aligned}
		\big(F_d^{abc}\big)_{ef} &= \big(F_{c^*}^{bad^*}\big)_{ef^*} = \big(F_{a^*}^{d^*cb}\big)_{e^*f} \\
		& = \sqrt{\frac{d_e d_f}{d_a d_c}} \big(F_{d^*}^{e^*bf^*}\big)_{a^*c^*}.
	\end{aligned}
	\label{eq:tetrahedral_symmetry}
\end{equation}
Tetrahedral symmetry is studied in detail in Ref.~\cite{Fuchs2023} and implies that~\cite[Sec.~8.1]{Wolf2020}
\begin{equation*}
	\begin{tikzpicture}[decoration={markings,mark=at position .5 with {\arrow[scale=1,thick,>=stealth]{>}}},baseline=(current bounding box.center)]
		\draw[{postaction=decorate}] (0.5,0) to node [above] {$e$} (-0.5,0);
		\draw[{postaction=decorate}] (-1,0.5) to node [right,midway] {$a$} (-0.5,0);
		\draw[{postaction=decorate}] (-1,-0.5) to node [right,midway] {$b$} (-0.5,0);
		\draw[{postaction=decorate}] (1,0.5) to node [left,midway] {$c$} (0.5,0);
		\draw[{postaction=decorate}] (1,-0.5) to node [left,midway] {$d$} (0.5,0);
	\end{tikzpicture}
	= \sum_f \big(F_d^{b^*a^*c^*}\big)_{ef}\;
	\begin{tikzpicture}[decoration={markings,mark=at position .5 with {\arrow[scale=1,thick,>=stealth]{>}}},baseline=(current bounding box.center)]
		\draw[{postaction=decorate}] (0,-0.5) to node[right, midway] {$f$} (0,0.5);
		\draw[{postaction=decorate}] (-1,0.5) to node[above] {$a$} (0,0.5);
		\draw[{postaction=decorate}] (1,0.5) to node[above] {$c$} (0,0.5);
		\draw[{postaction=decorate}] (-1,-0.5) to node[below] {$b$} (0,-0.5);
		\draw[{postaction=decorate}] (1,-0.5) to node[below] {$d$} (0,-0.5);
	\end{tikzpicture}\;.
\end{equation*}
In particular, it gives sense to drawing horizontal lines in string diagrams. The name tetrahedral symmetry stems from the evaluation of a string diagram in the form of a tetrahedron by applying a sequence of $F$-symbols. This can be done in four different ways, which yield the same result if and only if Eq.~\eqref{eq:tetrahedral_symmetry} is satisfied, see Ref.~\cite[Sec.~8.1]{Wolf2020}\@. Therefore, the tetrahedral symmetry relation is the categorical analogue of the closure constraint~\eqref{eq:closure_constraint} for spin networks.
	
Due to the braiding structure in $\C$, we can define the $R$-matrices
\begin{equation}
	\begin{tikzpicture}[decoration={markings,mark=at position 0.8 with {\arrow[scale=1,thick,>=stealth]{>}}},baseline=(current bounding box.center)] 
		\node (Y) at (1,0.8)[right] {$b$};
		\node (X) at (0,0.8)[left] {$a$};
		\draw[postaction={decorate}] (1,0) -- (0,1);
		\draw[overdraw=10pt,postaction={decorate}] (0,0) -- (1,1);
		\draw (0.5,-0.5) -- (1,0);
		\draw (0.5,-0.5) -- (0,0);
		\draw[postaction={decorate}] (0.5,-1) to node [left] {$c$} (0.5,-0.5);
	\end{tikzpicture}
	= R^{ab}_c
	\begin{tikzpicture}[decoration={markings,mark=at position 0.5 with {\arrow[scale=1,thick,>=stealth]{>}}},baseline=(current bounding box.center)] 
		\draw[{postaction=decorate}] (0,-1) to node [left,midway] {$c$} (0,0);
		\draw[{postaction=decorate}] (0,0) to node [below right,midway] {$b$} (0.75,0.75);
		\draw[{postaction=decorate}] (0,0) to node [below left,midway] {$a$} (-0.75,0.75);
	\end{tikzpicture}\;\label{eq:R-matrix}
\end{equation}
and their mirrored version
\begin{equation*}
	\begin{tikzpicture}[decoration={markings,mark=at position 0.8 with {\arrow[scale=1,thick,>=stealth]{<}}},baseline=(current bounding box.center),rotate=180] 
		\node (Y) at (1,0.8)[left] {$a$};
		\node (X) at (0,0.8)[right] {$b$};
		\draw[postaction={decorate}] (0,0) -- (1,1);
		\draw[overdraw=10pt,postaction={decorate}] (1,0) -- (0,1);
		\draw (0.5,-0.5) -- (1,0);
		\draw (0.5,-0.5) -- (0,0);
		\draw[postaction={decorate}] (0.5,-1) to node [left] {$c$} (0.5,-0.5);
	\end{tikzpicture}
	= \big(R^{ab}_c\big)^{-1}
	\begin{tikzpicture}[decoration={markings,mark=at position 0.5 with {\arrow[scale=1,thick,>=stealth]{>}}},baseline=(current bounding box.center),rotate=180] 
		\draw[{postaction=decorate}] (0,0) to node [right,midway] {$c$} (0,-1);
		\draw[{postaction=decorate}] (0.75,0.75) to node [above left,midway] {$a$} (0,0);
		\draw[{postaction=decorate}] (-0.75,0.75) to node [above right,midway] {$b$} (0,0);
	\end{tikzpicture}\;.
\end{equation*}
If $\C$ is unitary, $R$ will be a unitary matrix.

\subsection{Derivation of the categorical volume operator} \label{subsec:Qvol.derivation}
To compute the matrix elements of the generalized volume operator, we follow the same steps as for the standard $\SU(2)$ volume operator in LQG~\cite{Rovelli2014}\@. That is, we consider a tetrahedron, but the corresponding four-valent intertwiners are not necessarily irreducible representations of $\SU(2)$. Instead, we treat them as morphisms in a premodular category $\C$,
\begin{equation*}
	\ket{\iota_\kappa}=
	(d_{j_1} d_{j_2} d_{j_3} d_{j_4})^{-1/4}\;
	\begin{tikzpicture}[decoration={markings,mark=at position 0.6 with {\arrow[scale=1,thick,>=stealth]{>}}},baseline=(current bounding box.center),scale=0.5]
		\node (k) at (0,-0.4) {$\kappa$};
		\draw[{postaction=decorate}] (1,0.6) -- (0,0);
		\draw[{postaction=decorate}] (1,0.6) -- (2,1.2);
		\node (j4) at (2,1.6) {$j_4$};
		\draw[{postaction=decorate}] (1,0.6) -- (0.6,1.2);
		\node (j3) at (0.6,1.6) {$j_3$};
		\draw[{postaction=decorate}] (0,0) -- (-1,0.6);
		\draw[{postaction=decorate}] (-1,0.6) -- (-2,1.2);
		\node (j1) at (-2,1.6) {$j_1$};
		\draw[{postaction=decorate}] (-1,0.6) -- (-0.6,1.2);
		\node (j2) at (-0.6,1.6) {$j_2$};
	\end{tikzpicture}\;.
\end{equation*}
With this approach, we can make sense of Eq.~\eqref{eq:volume_derivation_starting_point} in terms of string diagrams. For this purpose, we project the diagram representing $\bra{\iota_\kappa} L_{13}^2 \ket{\iota_{\kappa'}}$ [cf.\ Eq.~\eqref{eq:volume_diagram}] to the plane.
We remark that when working with objects from categories in 3D (as for the quantum volume operator) it is \emph{necessary} to define the operator acting on those degrees of freedom by projecting onto a two-dimensional plane.
For instance, this has also been done in the context of Walker--Wang models~\cite{Walker2012}, which realize topological phases of matter in 3D using objects from unitary braided fusion categories. These are fundamentally different from their 2D Levin--Wen cousins~\cite{LW04} since some twist factors with $R$-matrices appear in the former. 
The projection onto the plane is achieved by using the completeness relation~\eqref{eq:completeness_relation}\@.
Furthermore, we use that
\begin{equation}
	L_{13}^2\;
	\begin{tikzpicture}[baseline=(current bounding box.center)]
		\node (j1_b) at (-0.5,-1.2) {$j_1$};
		\draw (-0.5,-1) -- (0,-0.5);
		\node (j3_b) at (0.5,-1.2) {$j_3$};
		\draw (0.5,-1) -- (0,-0.5);
		\draw (0,-0.5) to node[right] {$\ell$} (0,0.5);
		\node (j1_a) at (-0.5,1.2) {$j_1$};
		\draw (0,0.5) -- (-0.5,1);
		\node (j3_a) at (0.5,1.2) {$j_3$};
		\draw (0,0.5) -- (0.5,1);
	\end{tikzpicture}
	= \alpha_\ell
	\begin{tikzpicture}[baseline=(current bounding box.center)]
		\node (j1_b) at (-0.5,-1.2) {$j_1$};
		\draw (-0.5,-1) -- (0,-0.5);
		\node (j3_b) at (0.5,-1.2) {$j_3$};
		\draw (0.5,-1) -- (0,-0.5);
		\draw (0,-0.5) to node[right] {$\ell$} (0,0.5);
		\node (j1_a) at (-0.5,1.2) {$j_1$};
		\draw (0,0.5) -- (-0.5,1);
		\node (j3_a) at (0.5,1.2) {$j_3$};
		\draw (0,0.5) -- (0.5,1);
	\end{tikzpicture}
	\label{eq:Casimir_second_order}
\end{equation}
and obtain
\begin{align}
	\tensor{Q}{_\kappa^{\kappa'}}
	=& -\frac{\rmi}{4} (\alpha_\kappa - \alpha_{\kappa'}) \frac{1}{\sqrt{d_{j_1} d_{j_2} d_{j_3} d_{j_4}}}\nonumber\\
	&\sum_\ell \sqrt{\frac{d_\ell}{d_{j_1}d_{j_3}}}\; \alpha_\ell\;
	\begin{tikzpicture}[decoration={markings,mark=at position 0.5 with {\arrow[scale=1,thick,>=stealth]{>}}},baseline=(current bounding box.center),scale=0.8]
		\coordinate[] (kb) at (0,0);
		\coordinate[] (j1) at (-2,2);
		\coordinate[] (j2) at (1,3);
		\coordinate[] (j3) at (-1,3);
		\coordinate[] (j4) at (2,2);
		\coordinate[] (j3j4kb) at (1,1);
		\coordinate[] (j1j2kb) at (-1,1);
		\coordinate[] (l) at (-1,4);
		\coordinate[] (j1o) at (-2,5);
		\coordinate[] (j1j2k) at (-1,6);
		\coordinate[] (k) at (0,7);
		\coordinate[] (j3j4k) at (1,6);
		\coordinate[] (j4o) at (2,5);
		\coordinate[] (j2o) at (1,4);
		\node (textkb) at (0,-0.4) {$\kappa'$};
		\draw[{postaction=decorate}] (j3j4kb) -- (kb);
		\draw (j3j4kb) -- (j4);
		\draw[decoration={markings,mark=at position 0.25 with {\arrow[scale=1,thick,>=stealth]{>}}},{postaction=decorate},{postaction=decorate}] (j3j4kb) -- (j3);
		\draw[{postaction=decorate},overdraw=10pt] (j1j2kb) -- (j2);
		\draw[{postaction=decorate}] (kb) -- (j1j2kb);
		\draw[{postaction=decorate}] (j1j2kb) to node[left] {$j_1$} (j1);
		\draw[{postaction=decorate}] (j1) -- (j3);
		\draw[{postaction=decorate}] (j3) to node[left,midway] {$\ell$} (l);
		\draw[{postaction=decorate}] (l) -- (j1o);
		\draw[{postaction=decorate}] (k) -- (j3j4k);
		\draw (j3j4k) -- (j4o);
		\draw[{postaction=decorate}] (j4) to node[right] {$j_4$} (j4o);
		\draw[decoration={markings,mark=at position 0.8 with {\arrow[scale=1,thick,>=stealth]{>}}},{postaction=decorate},{postaction=decorate}] (l) -- (j3j4k);
		\draw[{postaction=decorate}] (j2) to node[left] {$j_2$} (j2o);
		\draw[{postaction=decorate},overdraw=10pt] (j2o) -- (j1j2k);
		\draw[{postaction=decorate}] (j1o) to node[left] {$j_1$} (j1j2k);
		\draw[{postaction=decorate}] (j1j2k) -- (k);
		\node (textkb) at (0,7.4) {$\kappa$};
		\node (textj3) at (0.7,1.8) {$j_3$};
		\node (jextj3o) at (0.7,5.3) {$j_3$};
	\end{tikzpicture}\;.\label{eq:Qkkb}
\end{align}
The diagram can be computed using the graphical calculus of string diagrams introduced in the previous section. This gives
\begin{widetext}
	\begingroup
	\allowdisplaybreaks
	\begin{align*}
		\begin{tikzpicture}[decoration={markings,mark=at position 0.5 with {\arrow[scale=1,thick,>=stealth]{>}}},baseline=(current bounding box.center),scale=0.8]
			\coordinate[] (kb) at (0,0);
			\coordinate[] (j1) at (-2,2);
			\coordinate[] (j2) at (1,3);
			\coordinate[] (j3) at (-1,3);
			\coordinate[] (j4) at (2,2);
			\coordinate[] (j3j4kb) at (1,1);
			\coordinate[] (j1j2kb) at (-1,1);
			\coordinate[] (l) at (-1,4);
			\coordinate[] (j1o) at (-2,5);
			\coordinate[] (j1j2k) at (-1,6);
			\coordinate[] (k) at (0,7);
			\coordinate[] (j3j4k) at (1,6);
			\coordinate[] (j4o) at (2,5);
			\coordinate[] (j2o) at (1,4);
			\node (textkb) at (0,-0.4) {$\kappa'$};
			\draw[{postaction=decorate}] (j3j4kb) -- (kb);
			\draw (j3j4kb) -- (j4);
			\draw[decoration={markings,mark=at position 0.25 with {\arrow[scale=1,thick,>=stealth]{>}}},{postaction=decorate},{postaction=decorate}] (j3j4kb) -- (j3);
			\draw[{postaction=decorate},overdraw=10pt] (j1j2kb) -- (j2);
			\draw[{postaction=decorate}] (kb) -- (j1j2kb);
			\draw[{postaction=decorate}] (j1j2kb) to node[left] {$j_1$} (j1);
			\draw[{postaction=decorate}] (j1) -- (j3);
			\draw[{postaction=decorate}] (j3) to node[left,midway] {$\ell$} (l);
			\draw[{postaction=decorate}] (l) -- (j1o);
			\draw[{postaction=decorate}] (k) -- (j3j4k);
			\draw (j3j4k) -- (j4o);
			\draw[{postaction=decorate}] (j4) to node[right] {$j_4$} (j4o);
			\draw[decoration={markings,mark=at position 0.8 with {\arrow[scale=1,thick,>=stealth]{>}}},{postaction=decorate},{postaction=decorate}] (l) -- (j3j4k);
			\draw[{postaction=decorate}] (j2) to node[left] {$j_2$} (j2o);
			\draw[{postaction=decorate},overdraw=10pt] (j2o) -- (j1j2k);
			\draw[{postaction=decorate}] (j1o) to node[left] {$j_1$} (j1j2k);
			\draw[{postaction=decorate}] (j1j2k) -- (k);
			\node (textkb) at (0,7.4) {$\kappa$};
			\node (textj3) at (0.7,1.8) {$j_3$};
			\node (jextj3o) at (0.7,5.3) {$j_3$};
		\end{tikzpicture}
		=&
		\big(F_0^{\kappa'j_3j_4}\big)^{-1}_{\kappa'^*j_4^*} \big(F_0^{\kappa j_3j_4}\big)^{\T}_{\kappa^*j_4^*}
		\begin{tikzpicture}[decoration={markings,mark=at position 0.45 with {\arrow[scale=1,thick,>=stealth]{>}}},baseline=(current bounding box.center),scale=0.8]
			\coordinate[] (kb) at (0,0);
			\coordinate[] (j1) at (-2,2);
			\coordinate[] (j2) at (1,3);
			\coordinate[] (lstart) at (-1,3);
			\coordinate[] (j4) at (2,2);
			\coordinate[] (j3) at (0.5,1.5);
			\coordinate[] (j1j2kb) at (-1,1);
			\coordinate[] (l) at (-1,4);
			\coordinate[] (j1o) at (-2,5);
			\coordinate[] (j1j2k) at (-1,6);
			\coordinate[] (k) at (0,7);
			\coordinate[] (j3o) at (0.5,5.5);
			\coordinate[] (j4o) at (2,5);
			\coordinate[] (j2o) at (1,4);
			\coordinate[] (j3j4k) at (-0.5,6.5);
			\coordinate[] (j3j4kb) at (-0.5,0.5);
			\draw (j4) -- (kb);
			\draw[decoration={markings,mark=at position 0.7 with {\arrow[scale=1,thick,>=stealth]{>}}},{postaction=decorate}] (j3) -- (lstart);
			\draw[{postaction=decorate}] (j3j4kb) -- (kb);
			\draw[{postaction=decorate},overdraw=10pt] (j1j2kb) -- (j2);
			\draw[{postaction=decorate}] (j3j4kb) to node[below left] {$\kappa'$} (j1j2kb);
			\draw[{postaction=decorate}] (j1j2kb) to node[left] {$j_1$} (j1);
			\draw[{postaction=decorate}] (j1) -- (lstart);
			\draw[{postaction=decorate}] (lstart) to node[left,midway] {$\ell$} (l);
			\draw[{postaction=decorate}] (l) -- (j1o);
			\draw (k) -- (j4o);
			\draw[{postaction=decorate}] (j4) to node[right] {$j_4$} (j4o);
			\draw[{postaction=decorate}] (l) -- (j3o);
			\draw[{postaction=decorate}] (j2) to node[left] {$j_2$} (j2o);
			\draw[{postaction=decorate},overdraw=10pt] (j2o) -- (j1j2k);
			\draw[{postaction=decorate}] (j1o) to node[left] {$j_1$} (j1j2k);
			\draw[{postaction=decorate}] (j1j2k) to node[above left] {$\kappa$} (j3j4k);
			\draw[{postaction=decorate}] (k) -- (j3j4k);
			\draw (j3o) -- (j3j4k);
			\draw (j3) -- (j3j4kb);
			\node (textj3) at (0.7,1.8) {$j_3$};
			\node (jextj3o) at (0.7,5.3) {$j_3$};
		\end{tikzpicture}\\
		=&
		\big(F_0^{\kappa'j_3j_4}\big)^{-1}_{\kappa'^*j_4^*} \big(F_0^{\kappa j_3j_4}\big)^{\T}_{\kappa^*j_4^*}
		\sum_{m,n} \big(F_{j_4^*}^{j_1j_2j_3}\big)_{\kappa'm} \left(\big(F_{j_4^*}^{j_1j_2j_3}\big)^\T_{\kappa n}\right)^{-1}
		\begin{tikzpicture}[decoration={markings,mark=at position 0.45 with {\arrow[scale=1,thick,>=stealth]{>}}},baseline=(current bounding box.center),scale=0.8]
			\coordinate[] (kb) at (0,0);
			\coordinate[] (j1) at (-2,2);
			\coordinate[] (j2) at (1,3);
			\coordinate[] (lstart) at (-1,3);
			\coordinate[] (j4) at (2,2);
			\coordinate[] (j3) at (0.5,1.5);
			\coordinate[] (l) at (-1,4);
			\coordinate[] (j1o) at (-2,5);
			\coordinate[] (k) at (0,7);
			\coordinate[] (j3o) at (0.5,5.5);
			\coordinate[] (j4o) at (2,5);
			\coordinate[] (j2o) at (1,4);
			\coordinate[] (j3j4k) at (-0.5,6.5);
			\coordinate[] (j3j4kb) at (-0.5,0.5);
			\coordinate[] (j2oo) at (-0.5,5.5);
			\coordinate[] (j2uu) at (-0.5,1.5);
			\coordinate[] (j2j3o) at (0,6);
			\coordinate[] (j2j3u) at (0,1);
			\draw (j4) -- (kb);
			\draw[decoration={markings,mark=at position 0.7 with {\arrow[scale=1,thick,>=stealth]{>}}},{postaction=decorate}] (j3) -- (lstart);
			\draw[{postaction=decorate}] (j3j4kb) -- (kb);
			\draw[{postaction=decorate},overdraw=10pt] (j2uu) -- (j2);
			\draw[{postaction=decorate}] (j3j4kb) to node[below left] {$j_1$} (j1);
			\draw[{postaction=decorate}] (j1) -- (lstart);
			\draw[{postaction=decorate}] (lstart) to node[left,midway] {$\ell$} (l);
			\draw[{postaction=decorate}] (l) -- (j1o);
			\draw (k) -- (j4o);
			\draw[{postaction=decorate}] (j4) to node[right] {$j_4$} (j4o);
			\draw[{postaction=decorate}] (l) -- (j3o);
			\draw[{postaction=decorate}] (j2) to node[left] {$j_2$} (j2o);
			\draw[{postaction=decorate},overdraw=10pt] (j2o) -- (j2oo);
			\draw[{postaction=decorate}] (j1o) to node[above left] {$j_1$} (j3j4k);
			\draw[{postaction=decorate}] (k) -- (j3j4k);
			\draw (j3o) -- (j2j3o);
			\draw[{postaction=decorate}] (j2j3o) to node[right] {$n$} (j3j4k);
			\draw (j3) -- (j2j3u);
			\draw[{postaction=decorate}] (j3j4kb) to node[right] {$m$} (j2j3u);
			\draw (j2oo) -- (j2j3o);
			\draw (j2uu) -- (j2j3u);
			\node (textj3) at (0.7,1.8) {$j_3$};
			\node (jextj3o) at (0.7,5.3) {$j_3$};
		\end{tikzpicture}\\
		=&
		\big(F_0^{\kappa'j_3j_4}\big)^{-1}_{\kappa'^*j_4^*} \big(F_0^{\kappa j_3j_4}\big)^{\T}_{\kappa^*j_4^*}
		\sum_{m,n} \big(F_{j_4^*}^{j_1j_2j_3}\big)_{\kappa'm} \left(\big(F_{j_4^*}^{j_1j_2j_3}\big)^\T_{\kappa n}\right)^{-1}
		R_m^{j_2j_3} \big(R_n^{j_3j_2}\big)^{-1}\\
		&\qquad\begin{tikzpicture}[decoration={markings,mark=at position 0.45 with {\arrow[scale=1,thick,>=stealth]{>}}},baseline=(current bounding box.center),scale=0.8]
			\coordinate[] (bottom) at (0,0);
			\coordinate[] (top) at (0,7);
			\coordinate[] (bottomleft) at (-2,2);
			\coordinate[] (bottomright) at (2,2);
			\coordinate[] (topleft) at (-2,5);
			\coordinate[] (topright) at (2,5);
			\coordinate[] (j1j4n) at (-1,6);
			\coordinate[] (j1j4m) at (-1,1);
			\coordinate[] (n) at (0,5);
			\coordinate[] (m) at (0,2);
			\coordinate[] (lstart) at (-2,3);
			\coordinate[] (lend) at (-2,4);
			\coordinate[] (j2start) at (0,3);
			\coordinate[] (j2end) at (0,4);
			\coordinate[] (tl) at (-1.5,4.5);
			\coordinate[] (tr) at (-0.5,4.5);
			\coordinate[] (bl) at (-1.5,2.5);
			\coordinate[] (br) at (-0.5,2.5);
			\draw[{postaction=decorate}] (j1j4m) -- (bottom);
			\draw[{postaction=decorate}] (j1j4m) to node[below left] {$j_1$} (bottomleft);
			\draw[{postaction=decorate}] (bottom) -- (bottomright);
			\draw[{postaction=decorate}] (topleft) to node[above left] {$j_1$} (j1j4n);
			\draw[{postaction=decorate}] (top) -- (j1j4n);
			\draw[{postaction=decorate}] (topright) -- (top);
			\draw[{postaction=decorate}] (bottomright) to node[right] {$j_4$} (topright);
			\draw[{postaction=decorate}] (n) to node[right] {$n$} (j1j4n);
			\draw[{postaction=decorate}] (j1j4m) to node[right] {$m$} (m);
			\draw[{postaction=decorate}] (lstart) to node[left] {$\ell$} (lend);
			\draw[{postaction=decorate}] (j2start) to node[right] {$j_2$} (j2end);
			\draw[{postaction=decorate}] (tr) to node[below] {$j_3$} (tl);
			\draw[{postaction=decorate}] (tl) -- (topleft);
			\draw[{postaction=decorate}] (tr) -- (n);
			\draw[{postaction=decorate}] (lend) -- (tl);
			\draw[{postaction=decorate}] (j2end) -- (tr);
			\draw[{postaction=decorate}] (bl) to node[below] {$j_3$} (br);
			\draw[{postaction=decorate}] (br) -- (j2start);
			\draw[{postaction=decorate}] (bl) -- (lstart);
			\draw[{postaction=decorate}] (bottomleft) -- (bl);
			\draw[{postaction=decorate}] (m) -- (br);
		\end{tikzpicture}\\
		=&
		\big(F_0^{\kappa'j_3j_4}\big)^{-1}_{\kappa'^*j_4^*} \big(F_0^{\kappa j_3j_4}\big)^{\T}_{\kappa^*j_4^*}
		\sum_{m,n,o,p} \big(F_{j_4^*}^{j_1j_2j_3}\big)_{\kappa'm} \left(\big(F_{j_4^*}^{j_1j_2j_3}\big)^\T_{\kappa n}\right)^{-1}
		R_m^{j_2j_3} \big(R_n^{j_3j_2}\big)^{-1}\\
		&\qquad \big(F_{j_2}^{\ell^*j_1n}\big)_{j_3o} \big(F_m^{j_1^*\ell j_2}\big)_{j_3^*p^*}\\
		&\qquad\begin{tikzpicture}[decoration={markings,mark=at position 0.45 with {\arrow[scale=1,thick,>=stealth]{>}}},baseline=(current bounding box.center),scale=0.8]
			\coordinate[] (bottom) at (0,0);
			\coordinate[] (top) at (0,7);
			\coordinate[] (bottomleft) at (-2,2);
			\coordinate[] (bottomright) at (2,2);
			\coordinate[] (topleft) at (-2,5);
			\coordinate[] (topright) at (2,5);
			\coordinate[] (j1j4n) at (-1,6);
			\coordinate[] (j1j4m) at (-1,1);
			\coordinate[] (n) at (0,5);
			\coordinate[] (m) at (0,2);
			\coordinate[] (lstart) at (-2,3);
			\coordinate[] (lend) at (-2,4);
			\coordinate[] (j2start) at (0,3);
			\coordinate[] (j2end) at (0,4);
			\coordinate[] (ctop) at (-1,4);
			\coordinate[] (cbottom) at (-1,3);
			\coordinate[] (tritop) at (-1,5);
			\coordinate[] (tribottom) at (-1,2);
			\draw[{postaction=decorate}] (j1j4m) -- (bottom);
			\draw[{postaction=decorate}] (j1j4m) to node[below left] {$j_1$} (bottomleft);
			\draw[{postaction=decorate}] (bottom) -- (bottomright);
			\draw[{postaction=decorate}] (topleft) to node[above left] {$j_1$} (j1j4n);
			\draw[{postaction=decorate}] (top) -- (j1j4n);
			\draw[{postaction=decorate}] (topright) -- (top);
			\draw[{postaction=decorate}] (bottomright) to node[right] {$j_4$} (topright);
			\draw[{postaction=decorate}] (n) to node[right] {$n$} (j1j4n);
			\draw[{postaction=decorate}] (j1j4m) to node[right] {$m$} (m);
			\draw[{postaction=decorate}] (lstart) to node[left] {$\ell$} (lend);
			\draw[{postaction=decorate}] (j2start) to node[right] {$j_2$} (j2end);
			\draw[{postaction=decorate}] (j2end) -- (ctop);
			\draw[{postaction=decorate}] (lend) -- (ctop);
			\draw[{postaction=decorate}] (cbottom) -- (j2start);
			\draw[{postaction=decorate}] (cbottom) -- (lstart);
			\draw[{postaction=decorate}] (bottomleft) -- (tribottom);
			\draw[{postaction=decorate}] (m) -- (tribottom);
			\draw[{postaction=decorate}] (tritop) -- (topleft);
			\draw[{postaction=decorate}] (tritop) -- (n);
			\draw[{postaction=decorate}] (cbottom) to node[left] {$p$} (tribottom);
			\draw[{postaction=decorate}] (ctop) to node[left] {$o$} (tritop);
		\end{tikzpicture}\\
		=&\big(F_0^{\kappa'j_3j_4}\big)^{-1}_{\kappa'^*j_4^*} \big(F_0^{\kappa j_3j_4}\big)^{\T}_{\kappa^*j_4^*}
		\sum_{m,n,o,p} \big(F_{j_4^*}^{j_1j_2j_3}\big)_{\kappa'm} \left(\big(F_{j_4^*}^{j_1j_2j_3}\big)^\T_{\kappa n}\right)^{-1}
		R_m^{j_2j_3} \big(R_n^{j_3j_2}\big)^{-1}\\
		&\qquad \big(F_{j_2}^{\ell^*j_1n}\big)_{j_3o} \big(F_m^{j_1^*\ell j_2}\big)_{j_3^*p^*}\sqrt{\frac{d_{j_1}d_m}{d_{j_4}}}\delta_{p^*,j_4^*}\sqrt{\frac{d_\ell d_{j_2}}{d_p}} \delta_{o,p^*} \sqrt{\frac{d_{j_1}d_n}{d_{j_4}}}\delta_{o,j_4^*}\;
		\begin{tikzpicture}[decoration={markings,mark=at position .5 with {\arrow[scale=1,thick,>=stealth]{<}}},baseline=(current bounding box.center)]
			\draw[{postaction=decorate}] (0,-0.5) to node [left,midway] {$j_4$} (0,0.5);
			\draw (0,0.5) -- (0.5,1) -- (1,0.5);
			\draw[{postaction=decorate}] (1,0.5) to node [right,midway] {$j_4$} (1,-0.5);
			\draw (1,-0.5) -- (0.5,-1) -- (0,-0.5);
		\end{tikzpicture}\\
		=&
		\big(F_0^{\kappa'j_3j_4}\big)^{-1}_{\kappa'^*j_4^*} \big(F_0^{\kappa j_3j_4}\big)_{j_4^*\kappa^*} d_{j_1} \sqrt{\frac{d_{j_2}d_\ell}{d_{j_4}}}
		\sum_{m,n} \sqrt{d_md_n} \big(F_{j_4^*}^{j_1j_2j_3}\big)_{\kappa'm} \big(F_{j_4^*}^{j_1j_2j_3}\big)^{-1}_{n\kappa}
		\\
		&\qquad R_m^{j_2j_3} \big(R_n^{j_3j_2}\big)^{-1}\big(F_{j_2}^{\ell^*j_1n}\big)_{j_3j_4^*} \big(F_m^{j_1^*\ell j_2}\big)_{j_3^*j_4^*},
	\end{align*}
	\endgroup
\end{widetext}
where the penultimate step uses the bigon relation~\eqref{eq:bigon_relation} three times. This expression can be further simplified using tetrahedral symmetry~\eqref{eq:tetrahedral_symmetry}\@. In particular, we can rewrite the following $F$-symbol by using the equality from the first to the second tetrahedral symmetry relation
\begin{equation}
	\big(F_m^{j_1^*\ell j_2}\big)_{j_3^*j_4^*} = \big(F_{j_2^*}^{\ell j_1^* m^*}\big)_{j_3^* j_4}.
\end{equation}
Then, we use the equality from the first to the fourth tetrahedral symmetry relation to rewrite
\begin{align}
	\big(F_{j_2}^{\ell^*j_1n}\big)_{j_3j_4^*} &= \sqrt{\frac{d_{j_3} d_{j_4}}{d_n d_\ell}} \big(F_{j_2^*}^{j_3^* j_1 j_4}\big)_{\ell n^*}\\
	\big(F_{j_2^*}^{\ell j_1^* m^*}\big)_{j_3^* j_4} &= \sqrt{\frac{d_{j_3} d_{j_4}}{d_m d_\ell}} \big(F_{j_2}^{j_3 j_1^* j_4^*}\big)_{\ell^* m}.
\end{align}
Inserting everything back into Eq.~\eqref{eq:Qkkb} finally gives
\begin{align}
	\tensor{Q}{_\kappa^{\kappa'}} =& -\frac{\rmi}{4} (\alpha_\kappa - \alpha_{\kappa'}) \big(F_0^{\kappa'j_3j_4}\big)^{-1}_{\kappa'^*j_4^*} \big(F_0^{\kappa j_3j_4}\big)_{j_4^*\kappa^*}\nonumber
	\\
	&\sum_{m,n} \big(F_{j_4^*}^{j_1j_2j_3}\big)_{\kappa'm} \big(F_{j_4^*}^{j_1j_2j_3}\big)^{-1}_{n\kappa} R_m^{j_2j_3} \big(R_n^{j_3j_2}\big)^{-1}\nonumber
	\\
	&\quad\sum_\ell \alpha_\ell \big(F_{j_2^*}^{j_3^* j_1 j_4}\big)_{\ell n^*} \big(F_{j_2}^{j_3 j_1^* j_4^*}\big)_{\ell^* m}.\label{eq:generalized_volume}
\end{align}
In the case where $\C$ is unitary, we have the additional relation
\begin{equation}
	\big(F_{d^*}^{a^*b^*c^*}\big)_{e^*f^*} = \overline{\big(F_d^{abc}\big)}_{ef}
\end{equation}
for the $F$-symbols. This allows us to further simplify Eq.~\eqref{eq:generalized_volume} for unitary categories. Here, we obtain
\begin{align}
	\tensor{Q}{_\kappa^{\kappa'}} =& -\frac{\rmi}{4} (\alpha_\kappa - \alpha_{\kappa'}) \overline{\big(F_0^{\kappa'j_3j_4}\big)}_{j_4^* \kappa'^*} \big(F_0^{\kappa j_3j_4}\big)_{j_4^*\kappa^*} \nonumber \\
	&\sum_{m,n} \big(F_{j_4^*}^{j_1j_2j_3}\big)_{\kappa'm} \overline{\big(F_{j_4^*}^{j_1j_2j_3}\big)}_{\kappa n} R_m^{j_2j_3} \overline{R_n^{j_2j_3}} \nonumber \\
	&\quad\sum_\ell \alpha_\ell \big(F_{j_2^*}^{j_3^* j_1 j_4}\big)_{\ell n^*} \overline{\big(F_{j_2^*}^{j_3^* j_1 j_4}\big)}_{\ell m^*}.\label{eq:generalized_volume_unitary}
\end{align}
Note that the summation indices in Eq.~\eqref{eq:generalized_volume} and Eq.~\eqref{eq:generalized_volume_unitary} range over the following fusion products:
\begin{align}
	\begin{aligned}
		m,n &\in (j_1\otimes j_4^*) \cap (j_2\otimes j_3)\\
		\ell &\in (j_1\otimes j_3) \cap (j_2\otimes j_4^*).
	\end{aligned}
\end{align}
Finally, we comment on some aspects of our construction starting with the quadratic Casimir operator $L_{13}$, whose eigenvalues appear in our formulas for the generalized volume operator. In Ref.~\cite{Links1992} it has been shown how to generalize the idea of a Casimir invariant in Lie groups to quasi-triangular Hopf algebras. Notice that ribbon Hopf algebras are quasi-triangular, so if $\C$ is the representation category of a Hopf algebra, then the theory of Ref.~\cite{Links1992} applies:
Let $A$ be a Hopf algebra and fix an invariant bilinear form $\langle\cdot,\cdot\rangle$ on $A$. Furthermore, let $M$ be a finite-dimensional subspace of $A$, which is stable under the adjoint action of elements in $A$ and on which $\langle\cdot,\cdot\rangle$ is nondegenerate. Such an $M$ is a submodule of $A$. If we choose a basis $\{a_i\}_{i=1}^m$ for $M$, then we can define the quadratic Casimir as
\begin{equation}
	C_M = \sum_{i=1}^m a_i(a^i),
	\label{eq:Casimir}
\end{equation}
where $\{a^i\}_{i=1}^m$ is the $\langle\cdot,\cdot\rangle$-dual basis. As per Eq.~\eqref{eq:Casimir_second_order}, $L_{13}^2$ is a second-order Casimir operator, i.e.\ $C_M\in A \otimes A$. Links and Gould prove that $C_M$ is in the centre of $A$~\cite[Prop.~1]{Links1992}\@. Furthermore, $C_M$ is proportional to the identity $I\otimes I\in A\otimes A$ due to Schur's Lemma. In our setting, $M$ is a simple object in $\C$ and the proportionality constant $C_M\propto I\otimes I$ is $\alpha_M$. Links and Gould construct $C_M$ for the cases where $A=\mathcal{D}(G)$~\cite[Sec.~4]{Links1992}, i.e.\ the quantum double of a finite group $G$, and for $A=U_q(\mathfrak{g})$~\cite[Sec.~5]{Links1992}, i.e.\ a quantum group associated to a simple Lie algebra $\mathfrak{g}$. For quantum groups, the $\alpha_M$ are explicitly constructed in~\cite[Sec.~5]{Links1992}\@. For quantum double models, they can be computed through the universal $R$-matrix in $A$. It is worth noting that the universal $R$-matrix differs from the categorical $R$-matrix introduced in Eq.~\eqref{eq:R-matrix}\@. While the categorical $R$-matrix concerns the braiding of two $A$-modules $M$, the universal $R$-matrix is an invertible element in $A\otimes A$ satisfying the Yang--Baxter equation. To compute $C_M$ for $A=\mathcal{D}(G)$, we must find the representation $R_M$ of the universal $R$-matrix in the submodule $M$. Then, for every $M$, the element $C_M=T(R_M) R_M$ defines the Casimir operator. Here, $T:A\otimes A\rightarrow A\otimes A, a\otimes b \mapsto b\otimes a$.
	
It is worth pointing out that in the case where $\alpha_\ell=1$ for all $\ell$, the volume operator becomes zero due to the term $(\alpha_\kappa - \alpha_{\kappa'})$. This is the case if $A$ is triangular so that $\C$ has a symmetric braiding.

Lastly, we would like to mention that our construction is compatible with defining curved tetrahedra, which have been considered in Ref.~\cite{Han2023}\@.
In this case, isotopy invariance~\eqref{eq:normalization_trivalent_vertices} ensures that the value of the tetrahedron does not change despite continuous deformation of its edges.
Therefore, our generalized quantum tetrahedron can be flatly embedded in an arbitrarily curved spacetime.

\subsection{Beyond ribbon fusion categories} \label{subsec:spherical.C}
In the derivation of the standard volume operator for $\SU(2)$, the closure constraint~\eqref{eq:closure_constraint} guarantees that any oriented triplet $\vec{L}_j,\vec{L}_k,\vec{L}_\ell$ in the definition of the volume operator~\eqref{eq:def_volume_operator} gives the same result. As mentioned previously, the category-theoretic analog of the closure constraint is the tetrahedral symmetry condition~\eqref{eq:tetrahedral_symmetry}\@. However, there is a subtlety, which comes from the fact that we have to project the diagram in Eq.~\eqref{eq:volume_diagram} to the plane to make it amenable to the string diagram calculus in fusion categories. In the absence of tetrahedral symmetry, different couplings $j_k$--$j_\ell$ are, \textit{a priori}, not equivalent topologically. For example, consider the case of coupling $j_2$ and $j_3$ in Eq.~\eqref{eq:volume_diagram}\@. Then, Eq.~\eqref{eq:volume_derivation_starting_point} becomes
\begin{equation}
	\tensor{\tilde{Q}}{_\kappa^{\kappa'}} = - \frac{\rmi}{4} (\alpha_\kappa - \alpha_{\kappa'}) \bra{\iota_\kappa} L_{23}^2 \ket{\iota_{\kappa'}}\label{eq:volume_derivation_starting_point_j23}
\end{equation}
where we have denoted the operator as $\tilde{Q}$ for distinction from $\hat{Q}$. The diagram corresponding to Eq.~\eqref{eq:volume_diagram} then becomes
\begin{equation}
	\begin{tikzpicture}[baseline=(current bounding box.center)]
		\draw[dashed] (0,0) arc (-40:40:1);
		\draw[rotate=180] (0,0) arc (40:-40:1);
		\draw (2,0) arc (-40:40:1);
		\draw[rotate=180] (-2,0) arc (40:-40:1);
		\draw (2,0) -- (1,-0.5);
		\draw (1,-0.5) -- (0,0);
		\draw (2,1.287) -- (1,1.8);
		\draw (1,1.8) -- (0,1.287);
		\node (kb) at (1,-0.8) {$\kappa'$};
		\node (k) at (1,2.1) {$\kappa$};
		\node (j1) at (-0.5,0.65) {$j_1$};
		\node (j4) at (2.5,0.65) {$j_4$};
		\node[draw,rectangle] (L13) at (1,0.65) {$L_{23}^2$};
		\draw[dashed] (0.225,0.65) -- (L13);
		\draw (1.775,0.65) -- (L13);
		\node (j3) at (1.6,1.15) {$j_3$};
		\node (j2) at (0.4,1.15) {$j_2$};
	\end{tikzpicture}\;.\label{eq:volume_diagram_j23}
\end{equation}
Now, projecting the diagram in Eq.~\eqref{eq:volume_diagram_j23} to the plane does not involve braiding anymore. More concretely, we obtain the following volume operator for this scenario
\begin{align}
	\tensor{\tilde{Q}}{_\kappa^{\kappa'}}
	=& -\frac{\rmi}{4} (\alpha_\kappa - \alpha_{\kappa'}) \frac{1}{\sqrt{d_{j_1} d_{j_2} d_{j_3} d_{j_4}}}\nonumber\\
	&\sum_\ell \sqrt{\frac{d_\ell}{d_{j_2}d_{j_3}}}\; \alpha_\ell\;
	\begin{tikzpicture}[decoration={markings,mark=at position 0.5 with {\arrow[scale=1,thick,>=stealth]{>}}},baseline=(current bounding box.center),scale=0.8]
		\coordinate[] (kb) at (0,0);
		\coordinate[] (j1) at (-2,2);
		\coordinate[] (j4) at (2,2);
		\coordinate[] (j3j4kb) at (1,1);
		\coordinate[] (j1j2kb) at (-1,1);
		\coordinate[] (j1o) at (-2,5);
		\coordinate[] (j1j2k) at (-1,6);
		\coordinate[] (k) at (0,7);
		\coordinate[] (j3j4k) at (1,6);
		\coordinate[] (j4o) at (2,5);
		\coordinate[] (j2j3) at (0,2);
		\coordinate[] (j2j3o) at (0,5);
		\node (textkb) at (0,-0.4) {$\kappa'$};
		\draw[{postaction=decorate}] (j3j4kb) -- (kb);
		\draw (j3j4kb) -- (j4);
		\draw[{postaction=decorate}] (kb) -- (j1j2kb);
		\draw (j1j2kb) -- (j1);
		\draw[{postaction=decorate}] (k) -- (j3j4k);
		\draw (j3j4k) -- (j4o);
		\draw[{postaction=decorate}] (j4) to node[right] {$j_4$} (j4o);
		\draw (j1o) -- (j1j2k);
		\draw[{postaction=decorate}] (j1j2k) -- (k);
		\draw[{postaction=decorate}] (j1) to node[left] {$j_1$} (j1o);
		\draw[{postaction=decorate}] (j1j2kb) to node[above left] {$j_2$} (j2j3);
		\draw[{postaction=decorate}] (j3j4kb) to node[above right] {$j_3$} (j2j3);
		\draw[{postaction=decorate}] (j2j3o) to node[below left] {$j_2$} (j1j2k);
		\draw[{postaction=decorate}] (j2j3o) to node[below right] {$j_3$} (j3j4k);
		\draw[{postaction=decorate}] (j2j3) to node[left] {$\ell$} (j2j3o);
		\node (textkb) at (0,7.4) {$\kappa$};
	\end{tikzpicture}\;.\label{eq:Qkkb_j23}
\end{align}
Due to the vector identities $\vec{v}\times\vec{w}=-\vec{w}\times\vec{v}$ and $\vec{w}\cdot(\vec{u}\times\vec{v})=\vec{u}\otimes(\vec{v}\times\vec{w})=\vec{v}\cdot(\vec{w}\times\vec{u})$ as well as rotational symmetry of the diagram \eqref{eq:volume_diagram}, all possible different couplings of $j_k$ and $j_\ell$ are equivalent to either $j_1$--$j_3$ or $j_2$--$j_3$. That is, the coupling of two neighbouring $j_{k}$--$j_{k+1}$ corresponds to $j_2$--$j_3$, whereas a coupling of two next-to-nearest neighbours $j_{k}$--$j_{k+2}$ corresponds to $j_1$--$j_3$ [subject to periodicity $j_{k+4}=j_k$]. The couplings $j_1$--$j_2$ and $j_3$--$j_4$ are topologically trivial and therefore forbidden. We can compute the diagram \eqref{eq:Qkkb_j23} using a sequence of $F$-symbols, namely
\begin{widetext}
	\begingroup
	\allowdisplaybreaks
	\begin{align*}
		\begin{tikzpicture}[decoration={markings,mark=at position 0.5 with {\arrow[scale=1,thick,>=stealth]{>}}},baseline=(current bounding box.center),scale=0.8]
			\coordinate[] (kb) at (0,0);
			\coordinate[] (j1) at (-2,2);
			\coordinate[] (j4) at (2,2);
			\coordinate[] (j3j4kb) at (1,1);
			\coordinate[] (j1j2kb) at (-1,1);
			\coordinate[] (j1o) at (-2,5);
			\coordinate[] (j1j2k) at (-1,6);
			\coordinate[] (k) at (0,7);
			\coordinate[] (j3j4k) at (1,6);
			\coordinate[] (j4o) at (2,5);
			\coordinate[] (j2j3) at (0,2);
			\coordinate[] (j2j3o) at (0,5);
			\node (textkb) at (0,-0.4) {$\kappa'$};
			\draw[{postaction=decorate}] (j3j4kb) -- (kb);
			\draw (j3j4kb) -- (j4);
			\draw[{postaction=decorate}] (kb) -- (j1j2kb);
			\draw (j1j2kb) -- (j1);
			\draw[{postaction=decorate}] (k) -- (j3j4k);
			\draw (j3j4k) -- (j4o);
			\draw[{postaction=decorate}] (j4) to node[right] {$j_4$} (j4o);
			\draw (j1o) -- (j1j2k);
			\draw[{postaction=decorate}] (j1j2k) -- (k);
			\draw[{postaction=decorate}] (j1) to node[left] {$j_1$} (j1o);
			\draw[{postaction=decorate}] (j1j2kb) to node[above left] {$j_2$} (j2j3);
			\draw[{postaction=decorate}] (j3j4kb) to node[above right] {$j_3$} (j2j3);
			\draw[{postaction=decorate}] (j2j3o) to node[below left] {$j_2$} (j1j2k);
			\draw[{postaction=decorate}] (j2j3o) to node[below right] {$j_3$} (j3j4k);
			\draw[{postaction=decorate}] (j2j3) to node[left] {$\ell$} (j2j3o);
			\node (textkb) at (0,7.4) {$\kappa$};
		\end{tikzpicture}
		=&
		\big(F_0^{j_1j_2\kappa'^*}\big)_{\kappa'j_1^*} \big(F_0^{j_1j_2\kappa^*}\big)_{\kappa j_1^*}^\T\;
		\begin{tikzpicture}[decoration={markings,mark=at position 0.5 with {\arrow[scale=1,thick,>=stealth]{>}}},baseline=(current bounding box.center),scale=0.8]
			\coordinate[] (kb) at (0,0);
			\coordinate[] (j1) at (-2,2);
			\coordinate[] (j4) at (2,2);
			\coordinate[] (j3j4kb) at (0.75,0.75);
			\coordinate[] (j1j2kb) at (-1,1);
			\coordinate[] (j1o) at (-2,5);
			\coordinate[] (j1j2k) at (-1,6);
			\coordinate[] (k) at (0,7);
			\coordinate[] (j3j4k) at (0.75,6.25);
			\coordinate[] (j4o) at (2,5);
			\coordinate[] (j2j3) at (0,1.5);
			\coordinate[] (j2j3o) at (0,5.5);
			\coordinate[] (j2j3k) at (1.25,5.75);
			\coordinate[] (lo) at (0,4.5);
			\coordinate[] (j2j3kb) at (1.25,1.25);
			\coordinate[] (l) at (0,2.5);
			\draw[{postaction=decorate}] (j3j4kb) -- (kb);
			\draw (j2j3kb) -- (j4);
			\draw (kb) -- (j1j2kb);
			\draw (j1j2kb) -- (j1);
			\draw[{postaction=decorate}] (k) -- (j2j3k);
			\draw (j2j3k) -- (j4o);
			\draw[{postaction=decorate}] (j4) to node[right] {$j_4$} (j4o);
			\draw (j1o) -- (j1j2k);
			\draw (j1j2k) -- (k);
			\draw[{postaction=decorate}] (j1) to node[left] {$j_1$} (j1o);
			\draw[{postaction=decorate}] (j3j4kb) -- (j2j3);
			\draw[{postaction=decorate}] (j2j3o) -- (j3j4k);
			\draw[{postaction=decorate}] (j2j3) to node[left] {$\ell$} (j2j3o);
			\draw[{postaction=decorate}] (lo) to node[below right] {$j_3$} (j2j3k);
			\draw[{postaction=decorate}] (j2j3k) to node[above right] {$\kappa$} (j3j4k);
			\draw[{postaction=decorate}] (j2j3kb) to node[above right] {$j_3$} (l);
			\draw[{postaction=decorate}] (j3j4kb) to node[below right] {$\kappa'$} (j2j3kb);
			\node (textj2) at (-0.25,1.5) {$j_2$};
			\node (textj2o) at (-0.25,5.5) {$j_2$};
		\end{tikzpicture}\\
		=&
		\big(F_0^{j_1j_2\kappa'^*}\big)_{\kappa'j_1^*} \big(F_0^{j_1j_2\kappa^*}\big)_{\kappa j_1^*}^\T
		\sum_{m,n} \big(F_{j_1^*}^{j_2j_3j_4}\big)_{\kappa'm}^{-1} \left(\big(F_{j_1^*}^{j_2j_3j_4}\big)_{\kappa n}^\T\right)^{-1}\;
		\begin{tikzpicture}[decoration={markings,mark=at position 0.5 with {\arrow[scale=1,thick,>=stealth]{>}}},baseline=(current bounding box.center),scale=0.8]
			\coordinate[] (kb) at (0,0);
			\coordinate[] (j1) at (-2,2);
			\coordinate[] (j4) at (2,2);
			\coordinate[] (j3j4kb) at (0.75,0.75);
			\coordinate[] (j1j2kb) at (-1,1);
			\coordinate[] (j1o) at (-2,5);
			\coordinate[] (j1j2k) at (-1,6);
			\coordinate[] (k) at (0,7);
			\coordinate[] (j3j4k) at (0.75,6.25);
			\coordinate[] (j4o) at (2,5);
			\coordinate[] (j2j3) at (0,1.5);
			\coordinate[] (j2j3o) at (0,5.5);
			\coordinate[] (n) at (0,5);
			\coordinate[] (m) at (0,2);
			\coordinate[] (l) at (0,3);
			\coordinate[] (lo) at (0,4);
			\coordinate[] (bul) at (-0.5,2.5);
			\coordinate[] (bur) at (0.5,2.5);
			\coordinate[] (bol) at (-0.5,4.5);
			\coordinate[] (bor) at (0.5,4.5);
			\draw[{postaction=decorate}] (j3j4kb) -- (kb);
			\draw (j3j4kb) -- (j4);
			\draw (kb) -- (j1j2kb);
			\draw (j1j2kb) -- (j1);
			\draw[{postaction=decorate}] (k) -- (j3j4k);
			\draw (j3j4k) -- (j4o);
			\draw[{postaction=decorate}] (j4) to node[right] {$j_4$} (j4o);
			\draw (j1o) -- (j1j2k);
			\draw (j1j2k) -- (k);
			\draw[{postaction=decorate}] (j1) to node[left] {$j_1$} (j1o);
			\draw (j3j4kb) -- (j2j3);
			\draw (j2j3o) -- (j3j4k);
			\draw[{postaction=decorate}] (n) to node[left] {$n$} (j2j3o);
			\draw[{postaction=decorate}] (j2j3) to node[left] {$m$} (m);
			\draw[{postaction=decorate}] (l) to node[left] {$\ell$} (lo);
			\draw (m) -- (bul);
			\draw[{postaction=decorate}] (bul) -- (l);
			\draw (m) -- (bur);
			\draw[{postaction=decorate}] (bur) -- (l);
			\draw (lo) -- (bol);
			\draw[{postaction=decorate}] (bol) -- (n);
			\draw (lo) -- (bor);
			\draw[{postaction=decorate}] (bor) -- (n);
			\node (textj2) at (-0.75,2.5) {$j_2$};
			\node (textj3) at (0.75,2.5) {$j_3$};
			\node (textj2o) at (-0.75,4.5) {$j_2$};
			\node (textj3o) at (0.75,4.5) {$j_3$};
		\end{tikzpicture}\\
		=&
		\big(F_0^{j_1j_2\kappa'^*}\big)_{\kappa'j_1^*} \big(F_0^{j_1j_2\kappa^*}\big)_{\kappa j_1^*}^\T
		\sum_{m,n} \big(F_{j_1^*}^{j_2j_3j_4}\big)_{\kappa'm}^{-1} \left(\big(F_{j_1^*}^{j_2j_3j_4}\big)_{\kappa n}^\T\right)^{-1}
		\frac{d_{j_2}d_{j_3}}{d_\ell} \delta_{m,\ell}\delta_{n,\ell}\\
		&
		\begin{tikzpicture}[decoration={markings,mark=at position 0.5 with {\arrow[scale=1,thick,>=stealth]{>}}},baseline=(current bounding box.center),scale=0.4]
			\coordinate[] (kb) at (0,0);
			\coordinate[] (j1) at (-2,2);
			\coordinate[] (j4) at (2,2);
			\coordinate[] (j3j4kb) at (1,1);
			\coordinate[] (j1j2kb) at (-1,1);
			\coordinate[] (j1o) at (-2,5);
			\coordinate[] (j1j2k) at (-1,6);
			\coordinate[] (k) at (0,7);
			\coordinate[] (j3j4k) at (1,6);
			\coordinate[] (j4o) at (2,5);
			\coordinate[] (j2j3) at (0,2);
			\coordinate[] (j2j3o) at (0,5);
			\draw[{postaction=decorate}] (j3j4kb) -- (kb);
			\draw (j3j4kb) -- (j4);
			\draw (kb) -- (j1j2kb);
			\draw (j1j2kb) -- (j1);
			\draw[{postaction=decorate}] (k) -- (j3j4k);
			\draw (j3j4k) -- (j4o);
			\draw[{postaction=decorate}] (j4) to node[right] {$j_4$} (j4o);
			\draw (j1o) -- (j1j2k);
			\draw (j1j2k) -- (k);
			\draw[{postaction=decorate}] (j1) to node[left] {$j_1$} (j1o);
			\draw (j3j4kb) -- (j2j3);
			\draw (j2j3o) -- (j3j4k);
			\draw[{postaction=decorate}] (j2j3) to node[left] {$\ell$} (j2j3o);
		\end{tikzpicture}\\
		=&
		\big(F_0^{j_1j_2\kappa'^*}\big)_{\kappa'j_1^*} \big(F_0^{j_1j_2\kappa^*}\big)_{\kappa j_1^*}^\T
		\big(F_{j_1^*}^{j_2j_3j_4}\big)_{\kappa'\ell}^{-1} \left(\big(F_{j_1^*}^{j_2j_3j_4}\big)_{\kappa \ell}^\T\right)^{-1}
		\frac{d_{j_2}d_{j_3}}{d_\ell} \sqrt{\frac{d_\ell d_{j_4}}{d_{j_1}}}
		\begin{tikzpicture}[decoration={markings,mark=at position .5 with {\arrow[scale=1,thick,>=stealth]{>}}},baseline=(current bounding box.center)]
			\draw[{postaction=decorate}] (0,-0.5) to node [left,midway] {$j_1$} (0,0.5);
			\draw (0,0.5) -- (0.5,1) -- (1,0.5);
			\draw[{postaction=decorate}] (1,0.5) to node [right,midway] {$j_1$} (1,-0.5);
			\draw (1,-0.5) -- (0.5,-1) -- (0,-0.5);
		\end{tikzpicture}\\
		=&
		\sqrt{\frac{d_{j_1}d_{j_4}}{d_\ell}} d_{j_2} d_{j_3} \big(F_0^{j_1j_2\kappa'^*}\big)_{\kappa'j_1^*} \big(F_0^{j_1j_2\kappa^*}\big)_{j_1^* \kappa}
		\big(F_{j_1^*}^{j_2j_3j_4}\big)_{\kappa'\ell}^{-1} \big(F_{j_1^*}^{j_2j_3j_4}\big)_{\ell \kappa}^{-1}.
	\end{align*}
	\endgroup	
\end{widetext}
In total, the volume operator for the $j_2$--$j_3$ coupling becomes
\begin{align}
	\tensor{\tilde{Q}}{_\kappa^{\kappa'}}=&-\frac{\rmi}{4} (\alpha_\kappa -\alpha_{\kappa'})\big(F_0^{j_1j_2\kappa'^*}\big)_{\kappa'j_1^*} \big(F_0^{j_1j_2\kappa^*}\big)_{j_1^* \kappa}\nonumber\\
	&\sum_{\ell} \alpha_\ell \big(F_{j_1^*}^{j_2j_3j_4}\big)_{\kappa'\ell}^{-1} \big(F_{j_1^*}^{j_2j_3j_4}\big)_{\ell \kappa}^{-1},\label{eq:volume_j2j3}
\end{align}
where the sum over $\ell$ goes over the fusion product
\begin{equation}
	\ell\in (j_2\otimes j_3)\cap (j_1^*\otimes j_4).
\end{equation}
In the unitary case, $\tensor{\tilde{Q}}{_\kappa^{\kappa'}}$ reduces to
\begin{align}
	\tensor{\tilde{Q}}{_\kappa^{\kappa'}}=&-\frac{\rmi}{4} (\alpha_\kappa -\alpha_{\kappa'})\big(F_0^{j_1j_2\kappa'^*}\big)_{\kappa'j_1^*} \overline{\big(F_0^{j_1j_2\kappa^*}\big)}_{\kappa j_1^*}\nonumber\\
	&\sum_{\ell} \alpha_\ell \overline{\big(F_{j_1^*}^{j_2j_3j_4}\big)}_{\ell\kappa'} \big(F_{j_1^*}^{j_2j_3j_4}\big)_{\ell\kappa}.\label{eq:volume_j2j3_unitary}
\end{align}
A remarkable property of the $j_2$--$j_3$ volume operator is that its derivation involves neither the $R$-matrices nor the tetrahedral symmetry condition. Therefore, we can use this construction to go beyond the case of ribbon fusion categories. In particular, Eq.~\eqref{eq:volume_j2j3} is well-defined for any spherical fusion category $\C$ and Eq.~\eqref{eq:volume_j2j3_unitary} requires $\C$ to be a unitary fusion category. Nevertheless, for ribbon fusion categories, tetrahedral symmetry ensures that both the $j_1$--$j_3$ volume operator $\hat{Q}$  [Eq.~\eqref{eq:generalized_volume_unitary}] and the $j_2$--$j_3$ volume operator $\tilde{Q}$[Eq.~\eqref{eq:volume_j2j3_unitary}] coincide up to a phase factor related to the braiding. For $\C=\SU(2)_k$, it can be verified that this phase factor is $(-1)^{\kappa+\kappa'+2j_1+2j_2}$.

\subsection{Hermiticity of the categorical volume operator} \label{subsec:Hermicity}
For the volume operator to be a physical observable, it must be Hermitian. By looking at a specific example, we will see that for non-unitary $\C$ the volume operator is \emph{not} Hermitian in general. This is not surprising, but rather the expected behavior of generic non-unitary models. The same observation has been made, for instance, in anyon chains~\cite{Ardonne2011} and Levin--Wen string-net models~\cite{Freedman2012, Hahn2020}\@. However, we can prove that the volume operator is always Hermitian if $\C$ is unitary.
Furthermore, even in the non-unitary case, the physical volume operator $\hat{V}$ as defined in Eq.~\eqref{eq:def:vol_op} is Hermitian by construction (as $\vert\hat{Q}\vert$ is positive).
	
Let us start by first considering the non-unitary case. For this, we will look at the standard example, which is the Yang--Lee category $\YL$. $\YL$ is the Galois conjugate of the Fibonacci category, which consists of the two integer (or half-integer) spin irreducible representations of $\SU(2)_3$. While the $\SU(2)_k$ categories are specified by the deformation parameter $q=\rme^{\frac{2\pi\rmi}{k+2}}$, the Galois conjugate categories have a deformation parameter $q'=\rme^{\frac{4\pi\rmi}{k+2}}$ (see Ref.~\cite{Freedman2012} for more details). $\YL$ has two simple objects, usually denoted by $1$ and $\tau$. Their respective quantum dimensions are $d_1=1$ and $d_\tau=\phi$, where $\phi=(1+\sqrt{5})/2$ is the golden ratio. $\YL$ has the fusion rules
\begin{equation}
	1\otimes a = a = a \otimes 1, \;\forall a\in I,\quad \tau\otimes\tau = 1\oplus\tau
\end{equation}
and non-trivial $F$-symbols
\begin{equation}
	F_\tau^{\tau\tau\tau}=\begin{pmatrix}
		-\phi & \rmi\sqrt{\phi}\\
		\rmi\sqrt{\phi} & \phi
	\end{pmatrix}.
\end{equation}
The $R$-matrix reads
\begin{equation}
	R=\begin{pmatrix}
		-\rme^{-\frac{3\pi\rmi}{5}} & 0 \\
		0 & -\rme^{-\frac{4\pi\rmi}{5}}
	\end{pmatrix},
\end{equation}
where the entries of $R$ correspond to the non-trivial exchange of two $\tau$ objects fusing to either $1$ (upper left) or $\tau$ (lower right), respectively. The Casimir eigenvalues are given by a $q'$-deformation of the standard $\SU(2)$ Casimir eigenvalues. That is, we define the $q'$-deformed integer
\begin{equation}
	\qddi{n}=\frac{q'^{\frac{n}{2}}-q'^{-\frac{n}{2}}}{q'^{\frac{1}{2}}-q'^{-\frac{1}{2}}},\quad n\in\mathbb{Z}_{\geq0},
\end{equation}
such that $\alpha_\ell =\qddi{\ell(\ell+1)}$. With this data, $\YL$ only admits one non-trivial volume operator, namely if $j_1=j_2=j_3=j_4=\tau$. This is
\begin{equation}
	Q=\begin{pmatrix}
		0 & c_1\\
		c_2 & 0
	\end{pmatrix},
\end{equation}
where
\begin{align*}
	c_1 =& -\frac{(-1)^{1/5} \left(-\phi +\sqrt{\phi }+(-1)^{1/5}\right) \phi ^{3/2}}{16 \left(1+(-1)^{1/5}+(-1)^{2/5}\right)^2}\\
	\approx& -0.013514 - 0.0038170\,\rmi\\
	c_2 =& \; \frac{1}{16} (-1)^{3/5} \left(1+(-1)^{4/5}\right)^2 \phi  (2 \phi -3)\\
	&\quad \left((-1)^{1/5} \phi ^2-\sqrt{\phi }\right)\\
	\approx& \; 0.010961 - 0.011673\,\rmi,
\end{align*}
which is clearly non-Hermitian.
	
To prove Hermitianity of the volume operator for unitary fusion categories $\C$, we have to show $\tensor{Q}{_\kappa^{\kappa'}}=\overline{\tensor{Q}{_{\kappa'}^{\kappa}}}$. To this end, we compute
\begin{align}
	\overline{\tensor{Q}{_{\kappa'}^{\kappa}}}
	=& \frac{\rmi}{4} (\overline{\alpha}_{\kappa'} - \overline{\alpha}_{\kappa}) \frac{1}{\sqrt{d_{j_1} d_{j_2} d_{j_3} d_{j_4}}}\nonumber\\
	&\sum_\ell \sqrt{\frac{d_\ell}{d_{j_1}d_{j_3}}}\; \overline{\alpha}_\ell\;
	\begin{tikzpicture}[decoration={markings,mark=at position 0.5 with {\arrow[scale=1,thick,>=stealth]{>}}},baseline=(current bounding box.center),scale=0.8]
		\coordinate[] (kb) at (0,0);
		\coordinate[] (j1) at (-2,2);
		\coordinate[] (j2) at (1,3);
		\coordinate[] (j3) at (-1,3);
		\coordinate[] (j4) at (2,2);
		\coordinate[] (j3j4kb) at (1,1);
		\coordinate[] (j1j2kb) at (-1,1);
		\coordinate[] (l) at (-1,4);
		\coordinate[] (j1o) at (-2,5);
		\coordinate[] (j1j2k) at (-1,6);
		\coordinate[] (k) at (0,7);
		\coordinate[] (j3j4k) at (1,6);
		\coordinate[] (j4o) at (2,5);
		\coordinate[] (j2o) at (1,4);
		\node (textkb) at (0,-0.4) {$\kappa'$};
		\draw[{postaction=decorate}] (j3j4kb) -- (kb);
		\draw (j3j4kb) -- (j4);
		\draw[decoration={markings,mark=at position 0.25 with {\arrow[scale=1,thick,>=stealth]{>}}},{postaction=decorate},{postaction=decorate}] (j3j4kb) -- (j3);
		\draw[{postaction=decorate},overdraw=10pt] (j1j2kb) -- (j2);
		\draw[{postaction=decorate}] (kb) -- (j1j2kb);
		\draw[{postaction=decorate}] (j1j2kb) to node[left] {$j_1$} (j1);
		\draw[{postaction=decorate}] (j1) -- (j3);
		\draw[{postaction=decorate}] (j3) to node[left,midway] {$\ell$} (l);
		\draw[{postaction=decorate}] (l) -- (j1o);
		\draw[{postaction=decorate}] (k) -- (j3j4k);
		\draw (j3j4k) -- (j4o);
		\draw[{postaction=decorate}] (j4) to node[right] {$j_4$} (j4o);
		\draw[decoration={markings,mark=at position 0.8 with {\arrow[scale=1,thick,>=stealth]{>}}},{postaction=decorate},{postaction=decorate}] (l) -- (j3j4k);
		\draw[{postaction=decorate}] (j2) to node[left] {$j_2$} (j2o);
		\draw[{postaction=decorate},overdraw=10pt] (j2o) -- (j1j2k);
		\draw[{postaction=decorate}] (j1o) to node[left] {$j_1$} (j1j2k);
		\draw[{postaction=decorate}] (j1j2k) -- (k);
		\node (textkb) at (0,7.4) {$\kappa$};
		\node (textj3) at (0.7,1.8) {$j_3$};
		\node (jextj3o) at (0.7,5.3) {$j_3$};
	\end{tikzpicture}\;.
\end{align}
We note that the string diagram in $\overline{\tensor{Q}{_{\kappa'}^{\kappa}}}$ is the same as the one for $\tensor{Q}{_\kappa^{\kappa'}}$ in Eq.~\eqref{eq:Qkkb} due to the mirror symmetry for unitary $\C$, i.e.\ taking the complex conjugate of a string diagram corresponds to flipping the diagram along the horizontal axis (while not flipping the arrows). Therefore, complex conjugation of the string diagram reverses the swap of the labels $\kappa,\kappa'$ in $\overline{\tensor{Q}{_{\kappa'}^{\kappa}}}$. Furthermore, we notice that by construction, all $\alpha_\ell\in\mathbb{R}$, see Eq.~\eqref{eq:Casimir}\@. Thus, we conclude that $\tensor{Q}{_\kappa^{\kappa'}}=\overline{\tensor{Q}{_{\kappa'}^{\kappa}}}$, which implies that $\hat{Q}$ is a Hermitian matrix. The same argumentation can be followed for the volume operator $\tilde{Q}$ that involves the $j_2$--$j_3$ coupling.

\section{Recovering the Standard $\SU(2)$ Volume Operator} \label{sec:recovery}
In this section, we show that our construction recovers the standard volume operator for $\SU(2)$, see Eq.~\eqref{eq:QRS} and Eq.~\eqref{eq:QCh}\@. For this, we consider the input category $\SU(2)_k$, which is the $q$-deformed version of $\SU(2)$ with deformation parameter $q=\rme^{\frac{2\pi\rmi}{k+2}}$. In the limit $k\rightarrow\infty$, we have $q \rightarrow 1$ and recover $\SU(2)$. The categorical data of $\SU(2)_k$ can be found in Ref.~\cite{Bonderson2007} or Ref.~\cite{Ardonne2010}\@. $\SU(2)_k$ has simple objects $I=\big\{\frac{1}{2},1,\frac{3}{2},\dots k\big\}$. With
\begin{equation}
	\qdi{n}=\frac{q^{\frac{n}{2}}-q^{-\frac{n}{2}}}{q^{\frac{1}{2}}-q^{-\frac{1}{2}}},\quad n\in\mathbb{Z}_{\geq0},
\end{equation}
being a $q$-deformed integer, we have $\alpha_\ell = \qdi{\ell(\ell+1)}$. In the limit $k\rightarrow\infty$, $\alpha_\ell\rightarrow\ell(\ell+1)$. The $R$-matrices read
\begin{align}
	R_j^{j_1 j_2} &= (-1)^{j-j_1-j_2} q^{\frac{1}{2}\big(j(j+1)-j_1(j_1+1)-j_2(j_2+1)\big)} \nonumber \\
	&\rightarrow (-1)^{j-j_1-j_2}
\end{align}
and the quantum dimensions become
\begin{equation}
	d_j = \frac{\sin\left(\frac{(2j+1)\pi}{k+2}\right)}{\sin\left(\frac{\pi}{k+2}\right)}
	\rightarrow 2j+1.
\end{equation}
In the same way, we evaluate the limit $k\rightarrow\infty$ for the $F$-symbols, which yields
\begin{align}
	\big(F_j^{j_1 j_2 j_3}\big)_{j_{12} j_{23}} \rightarrow & (-1)^{j_1+j_2+j_3+j} \sqrt{(2j_{12}+1)(2j_{23}+1)}\nonumber \\&\sixj{j_1}{j_2}{j_{12}}{j_3}{j}{j_{23}}.
\end{align}
Note that the $F$-symbols for $\SU(2)_k$ are always real for all $k$ and, in addition, all simple objects are self-dual, i.e.\ $j=j^*$ for all $j\in I$. The same properties also hold for $\SU(2)$. Since $\SU(2)_k$ is unitary, we can take the limit $k\rightarrow\infty$ of Eq.~\eqref{eq:generalized_volume_unitary}\@. This gives
\begin{align}
	\tensor{Q}{_\kappa^{\kappa'}}\rightarrow & -\frac{\rmi}{4} [\kappa(\kappa+1)-\kappa'(\kappa'+1)] (-1)^{\kappa+\kappa'} (-1)^{2(j_4-j_2)}\nonumber\\
	&(2j_4+1)(2\kappa+1)(2\kappa'+1) \sixj{\kappa'}{j_3}{j_4}{j_4}{0}{\kappa'}\nonumber\\
	&\sixj{\kappa}{j_3}{j_4}{j_4}{0}{\kappa} \sum_{m,n\in\mathcal{L}_{23}} (-1)^{m+n} (2m+1) (2n+1)\nonumber\\
	&\quad \sixj{j_1}{j_2}{\kappa'}{j_3}{j_4}{m} \sixj{j_1}{j_2}{\kappa}{j_3}{j_4}{n}\sum_{\ell\in\mathcal{L}_{13}} \ell(\ell+1)(2\ell+1)\nonumber\\
	&\qquad \sixj{j_3}{j_1}{\ell}{j_4}{j_2}{m} \sixj{j_3}{j_1}{\ell}{j_4}{j_2}{n},
\end{align}
where $\mathcal{L}_{13}=\{\vert j_1-j_3\vert,\dots,j_1+j_3\}\cap\{\vert j_2-j_4\vert,\dots,j_2+j_4\}$. Now, we use the following identity for $6j$-symbols involving a zero charge~\cite[Ch.~9]{Varshalovich1988}:
\begin{equation}
	\sixj{b}{a}{c}{c}{0}{b} = \sixj{a}{b}{c}{0}{c}{b} = \frac{(-1)^{a+b+c}}{\sqrt{(2b+1)(2c+1)}},
\end{equation}
where the first step follows from the fact that the $6j$-symbols are invariant under permutation of their columns. This leads to
\begin{align*}
	\tensor{Q}{_\kappa^{\kappa'}}\rightarrow & -\frac{\rmi}{4} [\kappa(\kappa+1)-\kappa'(\kappa'+1)] (-1)^{\kappa+\kappa'} (-1)^{2(j_3-j_2)}\\
	&\sqrt{(2\kappa+1)(2\kappa'+1)} \sum_{\ell\in\mathcal{L}_{13}} \ell(\ell+1)(2\ell+1) (-1)^{-2\ell}\\
	&\sum_{m\in\mathcal{L}_{23}} (-1)^{m+\ell+\kappa'} (2m+1) \sixj{j_3}{j_2}{m}{j_1}{j_4}{\kappa'} \sixj{j_1}{j_4}{m}{j_2}{j_3}{\ell}\\
	&\sum_{n\in\mathcal{L}_{23}} (-1)^{n+\ell+\kappa} (2n+1) \sixj{j_3}{j_2}{n}{j_1}{j_4}{\kappa} \sixj{j_1}{j_4}{n}{j_2}{j_3}{\ell},
\end{align*}
where we again used invariance under permutation of columns and also that the $6j$-symbols are invariant under simultaneous exchange of upper and lower labels in two columns. By inspection of the set $\mathcal{L}$, we notice that the factor $(-1)^{2\ell}$ can be rewritten as $(-1)^{2\ell}=(-1)^{2(j_2+j_4)}$. The next step is to use another identity of the $6j$-symbols, namely~\cite[Ch.~9]{Varshalovich1988}
\begin{equation*}
	\sum_x (-1)^{f+g+x} (2x+1) \sixj{a}{b}{x}{c}{d}{f} \sixj{c}{d}{x}{b}{a}{g} = \sixj{a}{d}{f}{b}{c}{g}.
\end{equation*}
Thus, we finally arrive at
\begin{align}
	\tensor{Q}{_\kappa^{\kappa'}}\rightarrow & -\frac{\rmi}{4} (-1)^{\kappa+\kappa'} (-1)^{2(j_3+j_4)} [\kappa(\kappa+1)-\kappa'(\kappa'+1)]\nonumber\\
	&\sqrt{(2\kappa+1)(2\kappa'+1)} \sum_{\ell\in\mathcal{L}_{13}} \ell(\ell+1)(2\ell+1)\nonumber\\
	&\sixj{j_3}{j_4}{\kappa'}{j_2}{j_1}{\ell} \sixj{j_3}{j_4}{\kappa}{j_2}{j_1}{\ell}.\label{eq:limit_j1-j3}
\end{align}
A similar calculation shows that the $j_2$--$j_3$ volume operator $\tilde{Q}$ has a limit
\begin{align}
	\tensor{\tilde{Q}}{_\kappa^{\kappa'}}\rightarrow & -\frac{\rmi}{4} (-1)^{2(\kappa+\kappa')} (-1)^{2(j_1+j_2+j_3+j_4)}\nonumber\\
	&[\kappa(\kappa+1)-\kappa'(\kappa'+1)]\sqrt{(2\kappa+1)(2\kappa'+1)}\nonumber\\
	&\sum_{\ell\in\mathcal{L}_{23}} \ell(\ell+1)(2\ell+1)\sixj{j_1}{j_2}{\kappa'}{j_3}{j_4}{\ell}
	\sixj{j_1}{j_2}{\kappa}{j_3}{j_4}{\ell}.\label{eq:limit_j2-j3}
\end{align}
Comparison with Eq.~\eqref{eq:QCh} shows that the $j_2$--$j_3$ volume operator $\tilde{Q}$~\eqref{eq:limit_j2-j3} coincides with the standard $\SU(2)$ volume operator~\eqref{eq:QCh} up to a phase factor of $(-1)^{2(\kappa+\kappa'+j_1+j_2+j_3+j_4)}$ in the limit $k\rightarrow\infty$. Likewise, the $j_1$--$j_3$ volume operator  $\hat{Q}$~\eqref{eq:limit_j1-j3} coincides with the standard $\SU(2)$ volume operator~\eqref{eq:QCh} up to a phase factor of $(-1)^{\kappa+\kappa'+2(j_3+j_4)}$ in the limit $k\rightarrow\infty$.

\section{Spectrum of the $\SU(2)_k$ Volume Operator} \label{sec:spectrum}
The volume operator is a physical observable quantifying the volume of the three-dimensional spatial component of discrete spacetime. Therefore, the main interest is its spectral properties. The spectral analysis of the $\SU(2)$ volume operator has been carried out in Refs.~\cite{Loll1996,BrunnemannThiemann2006,Meissner2006,BrunnemannRideout2008a,BrunnemannRideout2008a}\@. By restricting the possible $\SU(2)$ irreducible representation labels of each tetrahedron to a value up to $k$, we obtain a categorical volume operator with input category $\SU(2)_k$. As shown in Sec.~\ref{sec:recovery}, this volume operator coincides with the $\SU(2)$ volume operator in the limit $k\rightarrow\infty$. In this section, we study the changes in the (tetrahedron) volume's spectrum due to the deformation $k$, while keeping the $j$ labels attached to the tetrahedron faces fixed. We also study how the spectrum varies with $k$ for different values of the fixed $j$'s. To this end, we fix two irreducible representation labels $j_1=j_2$ and vary $j_3=j_4=j_\mathrm{max}$. This is analogous to the analysis carried out in Ref.~\cite{BrunnemannRideout2008a}\@. We remark that in the context of loop quantum gravity, fixing the values of the $j$ labels corresponds to fixing the areas of the tetrahedron's faces (as measured by the $q$-deformed area operator \cite{Smolinqdeformed}\@).
	
\begin{figure}
	\centering
	\includegraphics{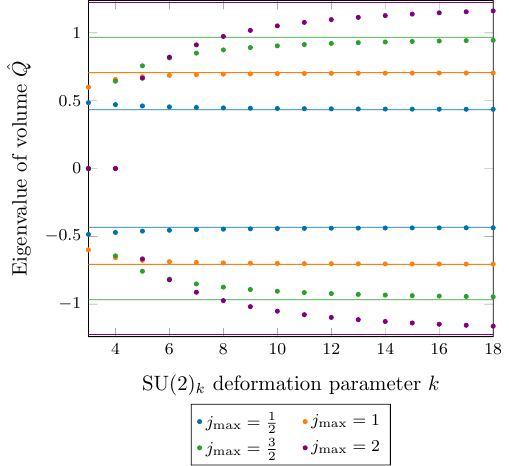}
	\caption{Eigenvalues of the quantum volume operator $\hat{Q}$ for $\SU(2)_k$ for different $k$. We fix the input labels to $j_1=j_2=\frac{1}{2}$ and show the eigenvalues of $\hat{Q}$ for different values of $j_3=j_4=j_\text{max}$ represented by the coloured dots. In addition, we plot the corresponding eigenvalues of the $\SU(2)$ volume operator (indicated by the constant horizontal lines), which coincide with the $\SU(2)_k$ eigenvalues in the limit $k\rightarrow\infty$.}
	\label{fig:j1/2}
\end{figure}

\begin{figure}
	\centering
	\includegraphics{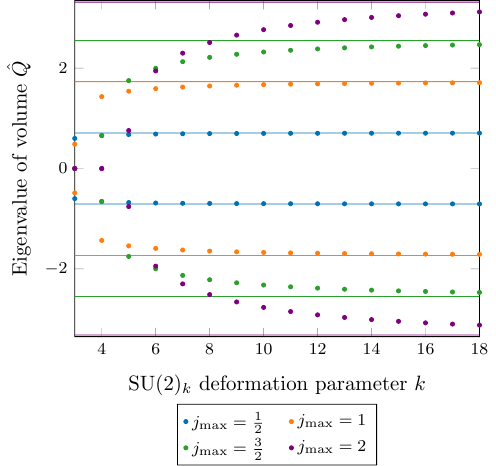}
	\caption{Eigenvalues of the quantum volume operator $\hat{Q}$ for $\SU(2)_k$ for different $k$. We fix the input labels to $j_1=j_2=1$ and show the nonzero eigenvalues of $\hat{Q}$ for different values of $j_3=j_4=j_\text{max}$ represented by the coloured dots. Except for the case where $j_\text{max}=\frac{1}{2}$, $\hat{Q}$ is a $3\times 3$ matrix and always admits an eigenvalue of zero. This behaviour is the same as for the $\SU(2)$ volume operator. The case $j_\text{max}=\frac{1}{2}$ is the same as as $j_\text{max}=1$ in Fig.~\ref{fig:j1/2}, cf.\ the orange dots and lines therein. In addition, we plot the corresponding eigenvalues of the $\SU(2)$ volume operator (indicated by the constant horizontal lines), which coincide with the $\SU(2)_k$ eigenvalues in the limit $k\rightarrow\infty$.}
	\label{fig:j1}
\end{figure}

For $j_1=j_2=\frac{1}{2}$, the volume operator becomes a $2\times2$ matrix, which we diagonalize. We find that the spectrum of the standard $\SU(2)$ volume operator and the generalized $\SU(2)_k$ volume operator are close to each other already for a moderately low value of $k$, see Fig.~\ref{fig:j1/2}\@. However, if $j_\mathrm{max}$ increases, the two spectra are significantly different for small $k$. In particular, the $\SU(2)_k$ volume operator admits smaller eigenvalues than the standard $\SU(2)$ volume operator. We find the same behaviour for the case where $j_1=j_2=1$, see Fig.~\ref{fig:j1}\@. This indicates that in spacetime regions with considerably different irreducible representation labels $j_1,j_2,j_3,j_4$ but relatively small maximum $j_\mathrm{max}$, the standard $\SU(2)$ volume operator tends to overestimate the volume. This might be due to weights in the volume from non-contributing higher irreducible representations.
	
\section{Discussion and Conclusions}\label{sec:discussion_conclusions}
In condensed matter physics, the categorification of spin models such as the Heisenberg model and certain lattice gauge theories has given rise to several physically rich anyonic lattice models that realize both conventional and exotic quantum phases of matter. Motivated by these developments, here we have introduced a categorification of the quantum volume operator [Sec.~\ref{sec:Qvol}] that plays a fundamental role in the description of discrete spacetime geometries underlying the formulation of background-independent candidate theories of quantum gravity such as LQG.

As discussed in the introduction, quantizing general relativity with a cosmological constant results in a $q$-deformation of LQG. However, a $q$-deformed volume operator has not been previously derived. In this work, we have derived a $q$-deformed volume operator using minimal physical assumptions, primarily driven by identifying and suitably generalizing the category-theoretical structures underlying the original construction of the SU(2) volume operator.

Subsequently, we have emphasized the physical relevance of various mathematical properties of category-theoretic objects and identified fusion categories --- a natural generalization of the representations of finite groups --- as the main ingredient in the construction. We have first constructed a categorified volume operator for ribbon fusion categories, which exhibit tetrahedral symmetry, [Subsec.~\ref{subsec:Qvol.derivation}] and then for more general spherical fusion categories without tetrahedral symmetry [Subsec.~\ref{subsec:spherical.C}]. We have shown that as long as the category is unitary, the corresponding generalized volume operator is guaranteed to be Hermitian [Subsec.~\ref{subsec:Hermicity}]. As an example, we consider the $q$-deformed group $\SU(2)_k$. After demonstrating that the standard $\SU(2)$ volume operator is recovered in the limit $k \to \infty$ [Sec.~\ref{sec:recovery}], we then compared the spectrum of the $q$-deformed $\SU(2)_k$ volume operator to that of its $\SU(2)$ counterpart [Sec.~\ref{sec:spectrum}].

An interesting direction for future research is to investigate the relationship between the $k$ parameter in our categorified volume operator and the cosmological constant in loop quantum gravity (LQG). Since our construction does not proceed from the quantization of classical gravity with the cosmological constant, it is unclear whether the $k$ (and thus $q$) parameter in our volume operator directly corresponds to the cosmological constant.

To establish a connection between $k$ and the cosmological constant, one could adopt an approach similar to that used in Ponzano-Regge geometries using a fixed triangulation of the spacetime manifold labeled by $SU(2)_k$ spins as discussed in Sec.\ref{sec:introduction}.
In the large spin limit, this resembles a modified Regge action that includes an explicit volume term, akin to the Einstein-Hilbert action with a cosmological constant (see e.g., Refs.~\cite{Smolinqdeformed, Smolinqdeformed1}). These calculations are performed in the small deformation limit, $k \gg 1$, which also regularizes the partition function. The partition function is proportional to a product over $6j$-symbols for all triangulation labellings and would diverge without a cutoff on the irreducible representations.

Our approach differs significantly as it is non-perturbative and uses a spin-network geometry instead of the Regge geometry, where the connection to an Einstein-Hilbert-like action is less clear. For instance, unlike in the Regge case, where the partition function sums over all admissible spin labelings of tetrahedra edges, a spin network also includes quantum degrees of freedom at the nodes (the intertwiners), which must also be considered.

\section*{Acknowledgements}
AH sincerely thanks Hanno Sahlmann for very interesting and helpful discussions and for pointing out several relevant references.
Furthermore, AH thanks Jacob C Bridgeman and Ramona Wolf for an interesting exchange. SM and GKB thank Sundance Bilson-Thompson for illuminating discussions.

AH acknowledges partial funding from the Sydney Quantum Academy.
SM is supported by the Quantum Gravity Unit of the Okinawa Institute of Science and Technology (OIST). G.K.B.\ acknowledges support from the Australian Research Council Centre of Excellence for Engineered Quantum Systems (Grant No. CE 170100009).
	
\bibliographystyle{prsty-title-hyperref}
\bibliography{LQG_Volume.bib}

\appendix
\section*{Appendix: Basic Concepts in Category Theory} \label{sec:category_theory}
\addcontentsline{toc}{section}{Appendix: Basic Concepts in Category Theory}
In this appendix, we introduce the basic category theoretical concepts that appear in the main text of this article. We will often not provide the full technical definitions, but instead focus on the implications of a categorical property to the graphical calculus of string diagrams. For an elementary physics-friendly introduction to category theory, we refer the reader to Refs.~\cite{Beer2018,Wolf2020}\@. A thorough and exhaustive mathematical introduction is provided in Ref.~\cite{Etingof2015}\@.
	
In the following, we denote a category with $\C$ and assume that it is small.\footnote{A category is called \emph{small} if both the class of objects and the class of morphisms are sets.} A category consists of objects and morphisms, which are maps between the objects. We denote the set of morphisms between objects $a,b\in\C$ as $\hom(a,b)$ and refer to it as hom-set. Graphically, we denote objects $a\in\C$ by oriented labeled strings
\begin{center}
	\begin{tikzpicture}[decoration={markings,mark=at position 0.5 with {\arrow[scale=1,thick,>=stealth]{>}}},baseline=(current bounding box.center)] 
		\node (end) at (0,1) {};
		\node (start) at (0,0) {};
		\draw[postaction={decorate}] (start) to node [right,midway] {$a$} (end);
	\end{tikzpicture}
\end{center}
and morphisms $f:a\rightarrow b$ between objects $a,b\in\C$ by circles with an input string $a$ and an output string $b$
\begin{center}
	\begin{tikzpicture}[decoration={markings,mark=at position .5 with {\arrow[scale=1,thick,>=stealth]{>}}},baseline=(current bounding box.center)]
		\node (f) at (0,0) [circle,draw] {$f$};
		\node (a) at (0,-1.25) {};
		\draw[{postaction=decorate}] (a) to node [right,midway] {$a$} (f);
		\node (b) at (0,1.25) {};
		\draw[{postaction=decorate}] (f) to node [right,midway] {$b$} (b);
	\end{tikzpicture}\;.	
\end{center}
The composition $g\circ f:a\rightarrow c$ of two morphisms $g : b \rightarrow c$ and $f : a \rightarrow b$ is given by stacking the circles and connecting the compatible strings
\begin{center}
	\begin{tikzpicture}[decoration={markings,mark=at position .5 with {\arrow[scale=1,thick,>=stealth]{>}}},baseline=(current bounding box.center)]
		\node (f) at (0,0) [circle,draw] {$f$};
		\node (a) at (0,-1.25) {};
		\draw[{postaction=decorate}] (a) to node [right,midway] {$a$} (f);
		\node (g) at (0,1.5) [circle,draw] {$g$};
		\draw[{postaction=decorate}] (f) to node [right,midway] {$b$} (g);
		\node (c) at (0,2.75) {};
		\draw[{postaction=decorate}] (g) to node [right,midway] {$c$} (c);
	\end{tikzpicture}\;.
\end{center}
The composition of morphisms is associative by definition.
	
We are interested in categories that carry a generalized tensor product structure. In category theoretical terms, this is called a \emph{monoidal structure} and comprises the following data:
\begin{enumerate}[(i)]
	\item a tensor product functor $\otimes:\C\times\C\rightarrow\C$;
	\item a unit object $0\in\C$;
	\item two natural isomorphisms called the left- and right-unitor such that $0\otimes a\simeq a \simeq a\otimes 0$ for all objects $a\in\C$;
	\item a natural isomorphism called the associator such that $(a\otimes b)\otimes c \simeq a\otimes (b\otimes c)$ for each triple of objects $a,b,c\in\C$.
\end{enumerate}
These structures satisfy certain consistency relations, see for instance Ref.~\cite[Def.~3.2]{Beer2018}\@. Most notably, the associators satisfy the pentagon identity. If $\C$ carries a monoidal structure, it is referred to as a \emph{monoidal category}. We remark that without loss of generality, we always work with \emph{skeletal} monoidal categories, see Ref.~\cite[Ch.~2.8]{Etingof2015}\@. This is standard in the physics literature. In terms of string diagrams, the tensor product $a\otimes b$ of two objects $a,b\in\C$ is denoted by writing the objects next to each other
\begin{center}
	\begin{tikzpicture}[decoration={markings,mark=at position 0.5 with {\arrow[scale=1,thick,>=stealth]{>}}},baseline=(current bounding box.center)] 
		\node (end_a) at (0,1) {};
		\node (start_a) at (0,0) {};
		\draw[postaction={decorate}] (start_a) to node [right,midway] {$a$} (end_a);
		\node (end_b) at (1,1) {};
		\node (start_b) at (1,0) {};
		\draw[postaction={decorate}] (start_b) to node [right,midway] {$b$} (end_b);
	\end{tikzpicture}\;.
\end{center}
	
The next concept we introduce is that of a \emph{braiding}. For a monoidal category $\C$, this is a family of natural isomorphisms $B_{a,b}:a\otimes b\rightarrow b\otimes a$. The $B_{a,b}$ satisfy certain consistency relations, which can be found for instance in Ref.~\cite[Def.~3.5]{Beer2018}\@. Furthermore, they satisfy the \emph{Yang--Baxter equation}. Graphically, the braiding $B_{a,b}$ is denoted by
\begin{center}
	\begin{tikzpicture}[decoration={markings,mark=at position 0.25 with {\arrow[scale=1,thick,>=stealth]{>}}},baseline=(current bounding box.center)] 
		\node (Y) at (1,0.2)[right] {$b$};
		\node (X) at (0,0.2)[left] {$a$};
		\draw[postaction={decorate}] (1,0) -- (0,2);
		\draw[overdraw=10pt,postaction={decorate}] (0,0) -- (1,2);
	\end{tikzpicture}\;
\end{center}
and its inverse $B^{-1}_{a,b}$ is
\begin{center}
	\begin{tikzpicture}[decoration={markings,mark=at position 0.25 with {\arrow[scale=1,thick,>=stealth]{>}}},baseline=(current bounding box.center)] 
		\node (Y) at (1,0.2)[right] {$b$};
		\node (X) at (0,0.2)[left] {$a$};
		\draw[postaction={decorate}] (0,0) -- (1,2);
		\draw[overdraw=10pt,postaction={decorate}] (1,0) -- (0,2);
	\end{tikzpicture}\;.
\end{center}
If the braiding is its own inverse, i.e.\ if $B_{a,b}=B_{b,a}^{-1}$, then we call the category \emph{symmetric}.
	
We will now assume that for all objects $a,b\in\C$ the hom-set $\hom(a,b)$ of morphisms from $a$ to $b$ is a vector space over the field $\k$. Furthermore, let the composition of morphisms in $\C$ be bilinear with respect to $\k$. A category with these properties is called \emph{$\k$-linear}. An object $a\in\C$ in a $\k$-linear category $\C$ is called \emph{simple} if $\hom(a,a)=\k$. We can then define the notion of a \emph{fusion category}, in which simple objects play a special role. That is, we assume $\C$ to be a monoidal $\k$-linear category and further assume the existence of a set $I$ of simple objects in $\C$. The set $I$ is assumed to satisfy the following three conditions:
\begin{enumerate}[(i)]
	\item the unit object is an element of $I$, i.e.\ $0\in I$;
	\item $\hom(a,b)=0$ for all simple objects $a,b\in I$ with $a\neq b$;
	\item any object $a\in\C$ can be written as a finite direct sum of simple objects, i.e.\ $a=\bigoplus_{i\in I} i$.
\end{enumerate}
Combining conditions (ii) and (iii) implies that all hom-sets $\hom(a,b)$ are finite-dimensional vector spaces and that for all $a,b\in\C$, $\hom(a,b)=\bigoplus_{i\in I} \hom(a,i)\otimes \hom(i,b)$~\cite[Ch.~4.4.1]{Turaev2017}\@. Due to this, we sometimes call the hom-sets \emph{fusion spaces}. In fusion categories, we can define the action of a fusion morphism $t:a\otimes b\rightarrow c$, which acts as
\begin{equation*}
	a\otimes b =\sum_{c\in I} N_{ab}^c\, c.
\end{equation*}
Here, the $N_{ab}^c$ are nonnegative integers known as \emph{fusion multiplicities}. They specify the number of different ways in which two objects $a,b$ may be fused to a third object $c$. We denote the hom-set of a fusion product $V_{ab}^c$, such that $\mathrm{dim}(V_{ab}^c)=N_{ab}^c$. Analogously, we can define a splitting morphism $t':c\rightarrow a\otimes b$ with corresponding splitting hom-set $V_{c}^{ab}$. In terms of string diagrams, fusion and splitting are represented by trivalent vertices
\begin{center}
	\begin{tikzpicture}[scale=1,decoration={markings,mark=at position .5 with {\arrow[thick,>=stealth]{>}}},baseline=(current bounding box.center)]
		\node (Xi) at (-0.25,0) {};
		\node (Xj) at (1.25,0) {};
		\node (Xk) at (0.5,2) {};
		\node (t1) at (0.5,1) [circle, draw] {$t$};
		\draw[{postaction=decorate}] (Xi) to node [above left,midway] {$a$} (t1);
		\draw[{postaction=decorate}] (Xj) to node [above right,midway] {$b$} (t1);
		\draw[{postaction=decorate}] (t1) to node [right,midway] {$c$} (Xk);
			
		\node (and) at (2.625,1) {and};
			
		\node (Xii) at (4,2) {};
		\node (Xjj) at (5.5,2) {};
		\node (Xkk) at (4.75,0) {};
		\node (t2) at (4.75,1) [circle, draw] {$t'$};
		\draw[{postaction=decorate}] (t2) to node [below left,midway] {$a$} (Xii);
		\draw[{postaction=decorate}] (t2) to node [below right,midway] {$b$} (Xjj);
		\draw[{postaction=decorate}] (Xkk) to node [right,midway] {$c$} (t2);.
	\end{tikzpicture}\,\,.
\end{center}
Since the hom-sets in fusion categories are vector spaces, we can fix a basis. For $V^{a_1,...,a_m}_{a'_1,...,a'_n}$, the standard basis is given by fusion diagrams of the form
\begin{equation}
	\begin{tikzpicture}[scale=1.5,baseline=(current bounding box.center), decoration={markings,mark=at position .6 with {\arrow[>=stealth]{>}}}]
		\coordinate (ydown) at (0,0);
		\coordinate (yup) at (0,0.5);
		\coordinate (t12) at (-0.5,1);
		\node (x1) at (-1,1.5) {$a_1$};
		\node (x2) at (0,1.5) {$a_2$};
		\node (xm) at (1,1.5) {$a_m$};
		\draw[{postaction=decorate}] (ydown) to node[right] {$b$} (yup);
		\draw[{postaction=decorate}] (yup) -- (xm);
		\draw[{postaction=decorate}] (yup) to node[below left] {$b_2$} (t12);
		\draw[{postaction=decorate}] (t12) -- (x2);
		\draw[{postaction=decorate}] (t12) -- (x1);
		\node (xsn) at (1,-1) {$a'_n$};
		\draw[{postaction=decorate}] (xsn) -- (ydown);
		\node (xs1) at (-1,-1) {$a'_1$};
		\node (xs2) at (0,-1) {$a'_2$};
		\coordinate (t12s) at (-0.5,-0.5);
		\draw[{postaction=decorate}] (xs1) -- (t12s);
		\draw[{postaction=decorate}] (xs2) -- (t12s);
		\draw[{postaction=decorate}] (t12s) to node[above left] {$b'_2$} (ydown);
		\node (downdots) at (0.5,-1) {$\cdots$};
		\node (updots) at (0.5,1.5) {$\cdots$};
		\node[rotate=45] (ysdots) at (-0.15,-0.35) {$\cdots$};
		\node[rotate=-45] (ydots) at (-0.15,0.85) {$\cdots$};
	\end{tikzpicture}.\label{eq:appendix_standard_basis}
\end{equation}
To simplify the notation, we do not indicate the fusion and splitting morphisms $t,t'$ explicitly here and will keep omitting them in what follows. In addition, we will also omit fusion multiplicities. However, all diagrams can be made general by adding a label at each trivalent vertex indicating the way the fusion was performed. By fixing the basis, the associators become matrices, called \emph{F-symbols} $F_d^{abc}:\hom(d,(a\otimes b)\otimes c)\rightarrow \hom(d,a\otimes(b\otimes c))$. They are represented graphically by
\begin{equation*}
	\begin{tikzpicture}[scale=1,baseline=(current bounding box.center), decoration={markings,mark=at position .6 with {\arrow[>=stealth]{>}}}]
		\node (u1) at (0,0) {$d$};
		\coordinate (v11) at (0,0.5);
		\coordinate (v12) at (-0.5,1);
		\node (x1) at (-1,1.5) {$a$};
		\node (y1) at (0,1.5) {$b$};
		\node (z1) at (1,1.5) {$c$};
			
		\draw[{postaction=decorate}] (u1) -- (v11);
		\draw[{postaction=decorate}] (v11) -- (z1);
		\draw[{postaction=decorate}] (v11) to node[below left] {$e$} (v12);
		\draw[{postaction=decorate}] (v12) -- (y1);
		\draw[{postaction=decorate}] (v12) -- (x1);
	\end{tikzpicture}
	=\sum_{f} \big(F_d^{abc}\big)_{ef}
	\begin{tikzpicture}[scale=1,baseline=(current bounding box.center), decoration={markings,mark=at position .6 with {\arrow[>=stealth]{>}}}]
		\node (u2) at (0,0) {$d$};
		\coordinate (v21) at (0,0.5);
		\coordinate (v22) at (0.5,1);
		\node (x2) at (-1,1.5) {$a$};
		\node (y2) at (0,1.5) {$b$};
		\node (z2) at (1,1.5) {$c$};
			
		\draw[{postaction=decorate}] (u2) -- (v21);
		\draw[{postaction=decorate}] (v21) -- (x2);
		\draw[{postaction=decorate}] (v21) to node[below right] {$f$} (v22);
		\draw[{postaction=decorate}] (v22) -- (y2);
		\draw[{postaction=decorate}] (v22) -- (z2);
	\end{tikzpicture}\;.
\end{equation*}
If all $F$-symbols are unitary, $\C$ is called a \emph{unitary} fusion category. For braided fusion categories, we can use the braiding to define the $R$-matrices
\begin{equation*}
	\begin{tikzpicture}[decoration={markings,mark=at position 0.8 with {\arrow[scale=1,thick,>=stealth]{>}}},baseline=(current bounding box.center)] 
		\node (Y) at (1,0.8)[right] {$a$};
		\node (X) at (0,0.8)[left] {$b$};
		\draw[postaction={decorate}] (1,0) -- (0,1);
		\draw[overdraw=10pt,postaction={decorate}] (0,0) -- (1,1);
		\draw (0.5,-0.5) -- (1,0);
		\draw (0.5,-0.5) -- (0,0);
		\draw[postaction={decorate}] (0.5,-1) to node [left] {$c$} (0.5,-0.5);
	\end{tikzpicture}
	= R^{ab}_c
	\begin{tikzpicture}[decoration={markings,mark=at position 0.5 with {\arrow[scale=1,thick,>=stealth]{>}}},baseline=(current bounding box.center)] 
		\draw[{postaction=decorate}] (0,-1) to node [left,midway] {$c$} (0,0);
		\draw[{postaction=decorate}] (0,0) to node [right,midway] {$a$} (0.75,0.75);
		\draw[{postaction=decorate}] (0,0) to node [left,midway] {$b$} (-0.75,0.75);
	\end{tikzpicture}\;.
\end{equation*}
They are also unitary matrices if the fusion category is unitary. In symmetric fusion categories, $R^\T$ is the inverse of $R$, where $^\T$ denotes transposition with respect to the standard basis.
	
Let us consider the special case where two objects fuse to the unit object $0\in\C$. This motivates the definition of a dual object. More precisely, for an object $a\in\C$ in a monoidal category $\C$, we can define its (left) dual object $a^*\in\C$ if there exist two morphisms $e_a: a^*\otimes a\rightarrow 0$ (coevaluation) and $i_a:0\rightarrow a\otimes a^*$ (evaluation) satisfying certain consistency relations~\cite[Sec.~3.3]{Beer2018}\@. The dual object is represented graphically by an arrow pointing downwards so that we have the relation
\begin{center}
	\begin{tikzpicture}[decoration={markings,mark=at position 0.5 with {\arrow[scale=1,thick,>=stealth]{>}}},baseline=(current bounding box.center)] 
		\node (end_a) at (0,1) {};
		\node (start_a) at (0,0) {};
		\draw[postaction={decorate}] (start_a) to node [left,midway] {$a^*$} (end_a);
		\node (eq) at (0.75,0.5) {$=$};
		\node (end_b) at (1.5,1) {};
		\node (start_b) at (1.5,0) {};
		\draw[postaction={decorate}] (end_b) to node [right,midway] {$a$} (start_b);
	\end{tikzpicture}\;.
\end{center}
The coevaluation and evaluation morphisms are depicted as
\begin{center}
	\begin{tikzpicture}[scale=1,decoration={markings,mark=at position .5 with {\arrow[thick,>=stealth]{>}}},baseline=(current bounding box.center)]
		\node (Xi) at (-0.25,0) {};
		\node (Xj) at (1.25,0) {};
		\node (Xk) at (0.5,2) {};
		\node (t1) at (0.5,1) [circle, draw] {$e_a$};
		\draw[{postaction=decorate}] (Xi) to node [above left,midway] {$a$} (t1);
		\draw[{postaction=decorate}] (Xj) to node [above right,midway] {$a^*$} (t1);
		\draw[dashed] (t1) to node [right,midway] {$0$} (Xk);
			
		\node (and) at (2.625,1) {and};
			
		\node (Xii) at (4,2) {};
		\node (Xjj) at (5.5,2) {};
		\node (Xkk) at (4.75,0) {};
		\node (t2) at (4.75,1) [circle, draw] {$i_a$};
		\draw[{postaction=decorate}] (t2) to node [below left,midway] {$a$} (Xii);
		\draw[{postaction=decorate}] (t2) to node [below right,midway] {$a^*$} (Xjj);
		\draw[dashed] (Xkk) to node [right,midway] {$0$} (t2);.
	\end{tikzpicture}\,\,.
\end{center}
Since the unit object $0\in\C$ is self-dual, we omit its arrow and draw it as a dashed line
\begin{center}
	\begin{tikzpicture}[decoration={markings,mark=at position 0.5 with {\arrow[scale=1,thick,>=stealth]{>}}},baseline=(current bounding box.center)] 
		\node (end) at (0,1) {};
		\node (start) at (0,0) {};
		\draw[dashed] (start) to node [right,midway] {$0$} (end);
	\end{tikzpicture}\;.
\end{center}
Right dual objects are defined analogously. A monoidal category is called \emph{rigid} (or sometimes \emph{compact}) if every object $a\in\C$ has both a left and a right dual object. In braided monoidal categories, the existence of a left dual already ensures the existence of a right dual~\cite[Prop.~7.2]{Joyal1993}\@. Notice that $e_a$ and $i_a$ are just special cases of fusion and splitting morphisms. The existence of coevaluation and evaluation morphisms allow the definition of a trace, see Ref.~\cite[Ch.~4.7]{Etingof2015}\@. For a morphism $f\in\hom(a,a^{**})$, the trace is represented graphically by
\begin{equation*}
	\mathrm{tr}(f)=
	\begin{tikzpicture}[decoration={markings,mark=at position .5 with {\arrow[scale=1,thick,>=stealth]{>}}},baseline=(current bounding box.center)]
		\node (f) at (0,0) [circle, draw] {$f$};
		\draw[{postaction=decorate}] (0,-1) to node [left,midway] {$a$} (f);
		\draw[{postaction=decorate}] (f) to node [left,midway] {$a^{**}$} (0,1);
		\draw (0,1) -- (0.5,1.5) -- (1,1);
		\draw[dashed] (0.5,1.5) to node [left,midway] {$0$} (0.5,2);
		\draw[{postaction=decorate}] (1,-1) to node [right,midway] {$a^*$} (1,1);
		\draw (0,-1) -- (0.5,-1.5) -- (1,-1);
		\draw[dashed] (0.5,-1.5) to node [left,midway] {$0$} (0.5,-2);
	\end{tikzpicture}\;.	
\end{equation*}
	
Now, let $\C$ be a rigid fusion category. If every object $a\in\C$ admits a natural isomorphism $p_a:a\simeq a^{**}$ called \emph{pivotal structure}, then $\C$ is called \emph{pivotal}. For an object $a\in\C$, the trace over the pivotal structure is called the \emph{quantum dimension}. We denote it by $d_a\equiv \mathrm{tr}(p_a)$ and use the simplified graphical notation
\begin{equation*}
	d_a=
	\begin{tikzpicture}[decoration={markings,mark=at position .5 with {\arrow[scale=1,thick,>=stealth]{>}}},baseline=(current bounding box.center)]
		\draw[{postaction=decorate}] (0,-0.5) to node [left,midway] {$a$} (0,0.5);
		\draw (0,0.5) -- (0.5,1) -- (1,0.5);
		\draw[{postaction=decorate}] (1,0.5) to node [right,midway] {$a$} (1,-0.5);
		\draw (1,-0.5) -- (0.5,-1) -- (0,-0.5);
	\end{tikzpicture}\;,
\end{equation*}
where we omit drawing the unit object. Notice that we have $d_a=d_{a^{**}}$ for all $a\in\C$~\cite[Ch.~4.7]{Etingof2015}\@. If, in addition, $d_a=d_{a^*}$ for all $a\in\C$, then $\C$ is called \emph{spherical}, see Ref.~\cite{Barrett1999}\@. Examples are unitary fusion categories, which are always spherical~\cite[Prop.~8.23]{Etingof2002}\@. Furthermore, every braided pivotal category is automatically spherical~\cite{Barrett1999}\@. In fact, the pivotal structure behaves particularly nicely in braided categories: Let $\C$ be a rigid braided monoidal category. Then $\C$ is called \emph{ribbon}~\cite{Reshetikhin1990} (or sometimes \emph{balanced}), if there exists a \emph{twist}. That is, a family of natural isomorphisms $\theta_a:a\rightarrow a$ for all $a\in\C$ graphically denoted as
\begin{center}
	\begin{tikzpicture}[decoration={markings,mark=at position .65 with {\arrow[scale=1,thick,>=stealth]{>}}},scale=.5,baseline=(current bounding box.center)]
		\node (x) at (0,0) {$a$};
		\draw[{postaction=decorate}] (x) -- (0,1);
		\begin{knot}[flip crossing/.list={4,5},consider self intersections=true,ignore endpoint intersections=false]
			\strand (0,1)to[out=90,in=180](1.25,2.1)to[out=0,in=90](1.75,1.5)to[out=270,in=0](1.25,.9)to[out=180,in=270](0,2);
			\strand (0,2) -- (0,2.5);
		\end{knot}
	\end{tikzpicture}\,\,.
\end{center}
It is easy to see that the pivotal structure defines a twist and vice versa. Hence, every rigid braided monoidal category is ribbon if and only if it is pivotal. In every ribbon fusion category, the $F$-symbols satisfy an algebraic symmetry property~\cite[Appendix~F]{Turaev2017}
\begin{align*}
	\big(F_d^{abc}\big)_{ef} &= \big(F_{c^*}^{bad^*}\big)_{ef^*} = \big(F_{a^*}^{d^*cb}\big)_{e^*f} \\
	&= \sqrt{\frac{d_e d_f}{d_a d_c}} \big(F_{d^*}^{e^*bf^*}\big)_{a^*c^*}
\end{align*}
called \emph{tetrahedral symmetry}~\cite{Fuchs2023}\@. This symmetry comes from evaluating a tetrahedral string diagram in different ways and imposing that all of them provide the same value~\cite[Sec.~8.1]{Wolf2020}\@. It can be shown by tetrahedral symmetry~\cite[Sec.~8.1]{Wolf2020} that the following graphical relation holds:
\begin{equation*}
	\begin{tikzpicture}[decoration={markings,mark=at position .5 with {\arrow[scale=1,thick,>=stealth]{>}}},baseline=(current bounding box.center)]
		\draw[{postaction=decorate}] (0.5,0) to node [above] {$e$} (-0.5,0);
		\draw[{postaction=decorate}] (-1,0.5) to node [right,midway] {$a$} (-0.5,0);
		\draw[{postaction=decorate}] (-1,-0.5) to node [right,midway] {$b$} (-0.5,0);
		\draw[{postaction=decorate}] (1,0.5) to node [left,midway] {$c$} (0.5,0);
		\draw[{postaction=decorate}] (1,-0.5) to node [left,midway] {$d$} (0.5,0);
	\end{tikzpicture}
	= \sum_f \big(F_d^{b^*a^*c^*}\big)_{ef}\;
	\begin{tikzpicture}[decoration={markings,mark=at position .5 with {\arrow[scale=1,thick,>=stealth]{>}}},baseline=(current bounding box.center)]
		\draw[{postaction=decorate}] (0,-0.5) to node[right, midway] {$f$} (0,0.5);
		\draw[{postaction=decorate}] (-1,0.5) to node[above] {$a$} (0,0.5);
		\draw[{postaction=decorate}] (1,0.5) to node[above] {$c$} (0,0.5);
		\draw[{postaction=decorate}] (-1,-0.5) to node[below] {$b$} (0,-0.5);
		\draw[{postaction=decorate}] (1,-0.5) to node[below] {$d$} (0,-0.5);
	\end{tikzpicture}\;.
\end{equation*}
To conclude, let us summarize the above discussion and refer back to the setting of the volume operator. Here, we assumed that $\C$ is a ribbon fusion category (also called \emph{premodular category)}. Notice that we necessarily need to work in a rigid braided monoidal category to ensure that the diagram defining the volume operator is well-defined. The above discussion shows that we only added two mild assumptions to this minimal requirement: First, we demand that $\C$ be a fusion category to guarantee that the $F$- and $R$-symbols are well-defined and to ensure that the volume operator has a matrix representation in the standard basis~\eqref{eq:appendix_standard_basis}\@. Second, we assume that $\C$ is pivotal. It is worth pointing out that it has been conjectured that every fusion category is pivotal~\cite{Etingof2002}, and there are currently no known examples of non-pivotal fusion categories. Therefore, this assumption is not restrictive. Nonetheless, due to the existence of a braiding, the pivotality of $\C$ implies sphericality as well as tetrahedral symmetry of the $F$-symbols.
	
\end{document}